%% file: sigconf.tex
\documentclass[acmtog,screen]{acmart} 
\AtBeginDocument{%
  }

\acmSubmissionID{318}

\setcctype{by}
\acmJournal{TOG}
\acmYear{2025} \acmVolume{44} \acmNumber{4} \acmArticle{} \acmMonth{8} \acmPrice{}\acmDOI{10.1145/3731154}
\acmISBN{978-1-4503-XXXX-X/18/06}

\input{preamble}

\begin{document}

\title{VR-Doh: Hands-on 3D Modeling in Virtual Reality}

\author{Zhaofeng Luo}
\authornote{Both authors contributed equally to this research.}
\affiliation{%
  \institution{Carnegie Mellon University}
  \country{USA}
}
\affiliation{%
  \institution{Peking University}
  \country{China}
}
\email{roushelfy@stu.pku.edu.cn}

\author{Zhitong Cui}
\authornotemark[1]
\affiliation{%
  \institution{Carnegie Mellon University}
  \country{USA}
}
\affiliation{%
  \institution{Zhejiang University}
  \country{China}
}
\email{zhitongcui@zju.edu.cn}

\author{Shijian Luo}
\affiliation{%
  \institution{Zhejiang University}
  \country{China}
}
\email{sjluo@zju.edu.cn}

\author{Mengyu Chu}
\affiliation{%
  \institution{Peking University, State Key Lab of General AI}
  \country{China}
}
\email{mchu@pku.edu.cn}

\author{Minchen Li}
\affiliation{%
  \institution{Carnegie Mellon University}
  \country{USA}
}
\email{minchernl@gmail.com}

\begin{abstract}
  We introduce VR-Doh, an open-source, hands-on 3D modeling system that enables intuitive creation and manipulation of elastoplastic objects in Virtual Reality (VR). By customizing the Material Point Method (MPM) for real-time simulation of hand-induced large deformations and enhancing 3D Gaussian Splatting for seamless rendering, VR-Doh provides an interactive and immersive 3D modeling experience. Users can naturally sculpt, deform, and edit objects through both contact- and gesture-based hand-object interactions.
  To achieve real-time performance, our system incorporates localized simulation techniques, particle-level collision handling, and the decoupling of physical and appearance representations, ensuring smooth and responsive interactions. VR-Doh supports both object creation and editing, enabling diverse modeling tasks such as designing food items, characters, and interlocking structures, all resulting in simulation-ready assets. User studies with both novice and experienced participants highlight the system's intuitive design, immersive feedback, and creative potential. Compared to existing geometric modeling tools, VR-Doh offers enhanced accessibility and natural interaction, making it a powerful tool for creative exploration in VR.
\end{abstract}





\begin{teaserfigure}
    \centering
    \noindent\includegraphics[width=\linewidth]{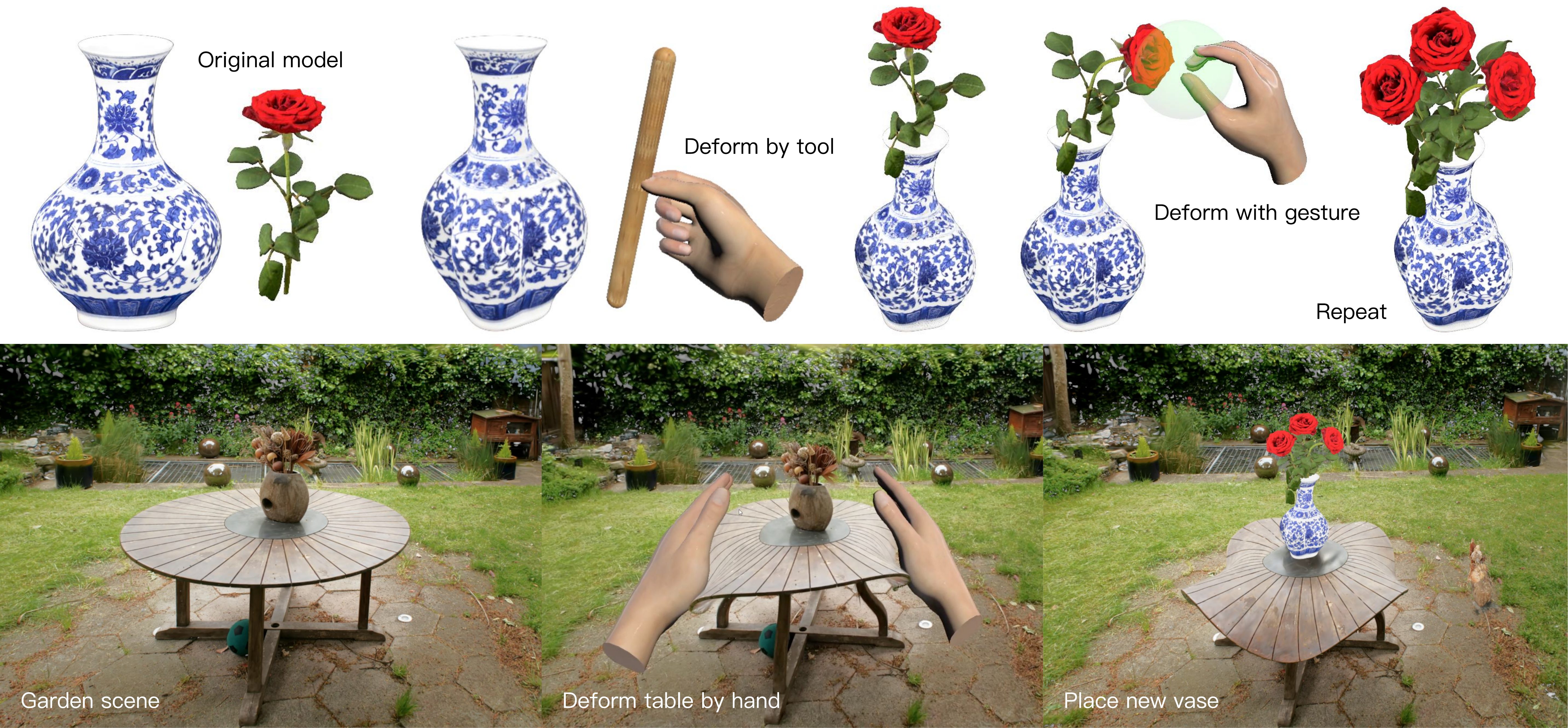}
    \caption{The figure demonstrates the intuitive and versatile capabilities of our VR-Doh system. In the upper row, users deform the vase and refine the roses with tools and gestures. The lower row showcases the system's versatility, enabling seamless table deformation by hand contact, object removal, and vase placement. This workflow simplifies complex editing tasks, delivering realistic results efficiently and intuitively.
    }
  \label{fig:1-teaser}
\end{teaserfigure}

\maketitle

\input{chapter01-introduction}
\input{chapter02-relatedwork}
\input{chapter03-design}
\input{chapter04-method}
\input{chapter05-evaluation}

\input{chapter06-case}

\input{chapter07-user}

\input{chapter08-discussion}
\input{chapter09-conclusion}

\begin{acks}
This work was supported in part by the Junior Faculty Startup Fund of Carnegie Mellon University. Z. Luo was partially supported by a summer internship scholarship from Peking University, and Z. Cui by a visiting scholarship from Zhejiang University. We thank the reviewers for their detailed and insightful feedback, and are especially grateful to all user study participants.
\end{acks}

\bibliographystyle{ACM-Reference-Format}
\bibliography{sigconf}

\clearpage
\includepdf[pages=-]{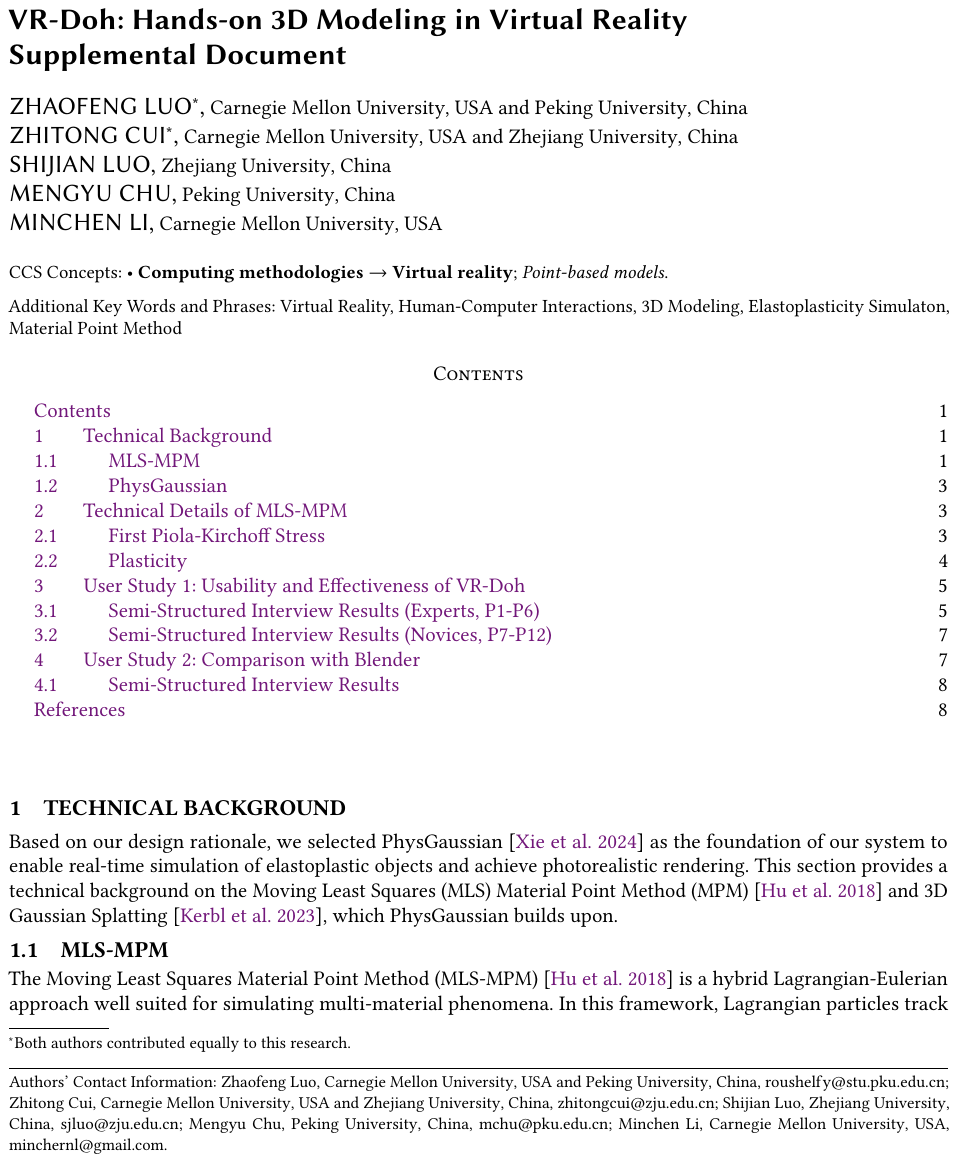}

\end{document}

%% file: preamble.tex
\usepackage[english]{babel}
\usepackage[utf8]{inputenc}
\usepackage{CJKutf8}

\newcommand{\COMMENTS}{yes}

\usepackage{color,balance,xspace,verbatim,ifthen,engord,unicode}

\usepackage{gensymb,amsmath,wasysym,marvosym}

\usepackage{newtxmath}

\usepackage{booktabs,colortbl,diagbox,multirow,tabularx,tablefootnote} 

\usepackage{dblfloatfix} 

\usepackage{caption}
\usepackage{subcaption} 
\usepackage{graphicx,epsfig,epstopdf,wrapfig}
\graphicspath{{./figures/}}
\DeclareGraphicsExtensions{.pdf,.mps,.png,.jpg,.eps,.PNG,.JPG}
\usepackage{algorithm,algpseudocode}

\usepackage{url}

\usepackage{pdfpages}

\newcommand{\figcaption}[1]{\vspace{-8mm}\caption{#1}\vspace{-4mm}}

\usepackage{tikz}

\newcommand*\circled[1]{\tikz[baseline=(char.base)]{\node[shape=circle,draw,inner sep=0.5pt] (char) {#1};}}

\usepackage{courier} 

\usepackage{enumitem}

\ifthenelse{\equal{\COMMENTS}{yes}}{
	\newcommand{\todo}[1]{\textcolor{red}{\textbf{TODO:} #1}}

	\newcommand{\fye}[1]{\textcolor{red}{#1}}  
	\newcommand{\remind}[1]{\footnote{\textit{\textcolor{red}{\textbf{Remind:} #1}}}}
	
	\newcommand{\del}[1]{\color{blue} {\sout{#1}}}
	\newcommand{\p}[1]{\vskip 1ex \noindent\colorbox{yellow}{\parbox{\columnwidth}{#1}}\vskip 4pt}
	\newcommand{\note}[1]{\vskip 4ex \noindent\colorbox{yellow}{\parbox{\columnwidth}{#1}}\vskip 6ex} 
	\newcommand{\q}[1]{\vskip 1ex \noindent\colorbox{magenta}{\parbox{\columnwidth}{\textbf{Question:} #1}}\vskip 4pt} 
	\newcommand{\qa}[1]{\hl{\textbf{Answer:} #1}}
}{
	\newcommand{\todo}[1]{}
	
	\newcommand{\fye}[1]{}
	\newcommand{\remind}[1]{}

	\newcommand{\del}[1]{}
	\newcommand{\p}[1]{}
	\newcommand{\note}[1]{}
\newif\ifnotnewacm 
\notnewacmtrue 

\newif\ifheadnice 
\headnicetrue 
	
	\newcommand{\q}[1]{}
	\newcommand{\qa}[1]{}	
} 
\newif\ifnotnewacm 
\notnewacmtrue 

\newif\ifheadnice 
\headnicetrue 

\ifheadnice
\makeatletter
\renewcommand{\section}{\@startsection{section}{1}{\z@}{-.8ex \@plus -.4ex \@minus -.4ex}{.8ex \@plus .4ex \@minus .4ex}{\normalfont\large\bfseries\MakeTextUppercase}} 
\renewcommand{\subsection}{\@startsection{subsection}{2}{\z@}{-.8ex\@plus -.2ex \@minus -.2ex}{.4ex \@plus .2ex \@minus .2ex}{\normalfont\large\rmfamily\bfseries}} 
\renewcommand{\subsubsection}{\@startsection{subsubsection}{2}{\z@}{-.4ex\@plus -.2ex \@minus -.2ex}{.2ex \@plus .2ex \@minus .2ex}{\normalfont\large\slshape}}

\makeatother
\else
\fi

%% file: chapter01-introduction.tex
\section{Introduction}

With the growing demand for digital content, there is an urgent need to address key challenges in 3D content creation: ease of use, scalability, efficiency, and quality. Traditional 3D modeling tools impose significant barriers for novice users, requiring expertise in real-world observation, specialized techniques, and iterative fine-tuning. This reliance on professional skills limits accessibility, hindering wider participation in high-quality content creation.

Meanwhile, 3D modeling in Virtual Reality (VR) has become increasingly popular due to its ability to provide immersion and depth, thereby enhancing designers' creativity and collaboration~\cite{rosales2019surfacebrush, yu2021cassie}. Commercial tools like Shapelab\footnote{\url{https://shapelabvr.com/}}
and Gravity Sketch\footnote{\url{https://gravitysketch.com/}}  have advanced this vision. However, their approaches are still limited to procedural methods, such as drawing curves to form surfaces from scratch, which are restricted to geometric techniques and require significant skills. 

In contrast, physics-aware shape modeling introduces a novel paradigm that aligns more closely with human intuition and natural interactions. In this approach, objects are simulated as deformable solids, reshaped through contact forces or boundary conditions specified by the user~\cite{Fang2021IDP}. In the context of VR, using hands as an input modality is particularly compelling, as it mimics the way users naturally manipulate physical objects, such as clay. This approach leverages users' prior experience with creating real-world objects, making the modeling process more intuitive and accessible~\cite{schulz2019rodmesh, arora2019magicalhands, pihuit2008hands}.

Building on these foundational advantages, we present \textit{VR-Doh}, an immersive VR modeling system that integrates physics-based simulation with natural hand interactions to achieve intuitive 3D content creation. Our system addresses two fundamental challenges: 1) enabling physically realistic material deformations through direct hand manipulation, and 2) maintaining real-time performance for responsive interaction despite complex physics computations. 

The foundation of VR-Doh lies in our customization of the Material Point Method (MPM) \cite{jiang2016material, hu2018moving}, a hybrid Lagrangian-Eulerian simulation framework with great potential for realistic, interactive modeling. We leverage MPM’s efficacy in handling large elastoplastic deformations to enable natural hand-object interactions, closely mimicking real-world clay manipulation. Furthermore, its particle-based formulation integrates naturally with modern scene representations such as 3D Gaussian Splatting (GS) \cite{kerbl20233d}, allowing VR-Doh to support not only freeform modeling from scratch but also editing of existing photorealistic 3D scenes via extensions of PhysGaussian~\cite{xie2024physgaussian}.
To address the challenges of standard MPM for real-time performance, we introduce three synergistic optimizations:
1) Particle-level collision handling resolves fine geometric details during intensive hand-object interactions without requiring high-resolution grids;
2) Decoupled physical and appearance representations allow detailed 3D GS rendering to be driven by lightweight simulation proxies; and
3) Localized simulation dynamically concentrates computational effort in regions of active manipulation, ensuring responsive performance. 
Complementing the technical innovations, we design an interaction toolkit that maps hand motions to a wide range of modeling operations. Users can control deformation tools, such as rods and planar slabs, for more structured edits, while mid-air gestures enable precise selection and manipulation via force fields. This multi-modal interaction paradigm broadens creative possibilities while keeping interactions grounded in real-world physics.

To evaluate VR-Doh’s usability and effectiveness, we conducted user studies with participants of varying expertise. The study highlights its intuitive design, immersive feedback, and potential for complementing traditional modeling software in creative exploration. These findings offer insights into how users naturally leverage physics-based interactions, paving the way for future improvements and more immersive 3D modeling experiences in VR.

Our key contributions are summarized as follows:
\begin{itemize}
    \item \textit{A novel physics-based VR 3D modeling paradigm}: We present VR-Doh, a VR system that integrates physics-based simulations to deliver intuitive and immersive workflows, making 3D modeling more accessible to users of varying expertise.
    \item \textit{Intuitive modeling through natural interactions}: VR-Doh uses hand-based contact and gesture inputs together with deformation tools to support tasks from basic shape editing to complex modeling, mimicking real-world interactions.
    \item \textit{Real-time performance through technical innovations}: We ensure smooth and responsive interactions with optimizations such as localized simulation, decoupled appearance and physical representations, and particle-level collision handling for efficient computation.
    \item \textit{Comprehensive evaluation}: Extensive user studies and experiments validate VR-Doh's effectiveness, offering insights into how physics-based hand-object interactions enhance intuitive and creative 3D modeling in VR.
\end{itemize}

A key advantage of our approach is its ability to make 3D modeling in VR significantly more intuitive and accessible, especially for novice users. By incorporating physics-based deformations, VR-Doh provides a realistic and immersive modeling experience that surpasses traditional geometric modeling approaches. Hands-on input lowers the barrier to entry, enabling users to interact with virtual objects naturally and intuitively, while flexible selection of operation areas improves efficiency and enhances immersion. This direct interaction mimics real-world manipulation, allowing users to predict outcomes based on everyday experiences and reducing the reliance on specialized skills and technical expertise. VR-Doh is open-sourced at \url{https://github.com/Simulation-Intelligence/VR-Doh}.

%% file: chapter02-relatedwork.tex
\section{Related Work}


\paragraph{Virtual Clay Modeling}
Virtual modeling of clay-like materials using dexterous input offers significant advantages over traditional 3D modeling, including high expressivity and a lower learning curve, effectively replicating the tactile experience of working with physical clay~\cite{sheng2006interface, chatterjee2024free}. Despite these benefits, the computational demands of simulating dexterous hand-object interactions have limited the integration of virtual clay modeling into VR environments, with most existing implementations constrained to 2D screens. Early approaches to virtual clay modeling utilized dynamic subdivision solids~\cite{mcdonnell2001virtual} and sculpting tools~\cite{galyean1991sculpting}, followed by the application of plasticity models to capture essential physical properties of real clay, such as mass conservation and surface tension effects~\cite{dewaele2004interactive}. To improve interactivity, \citet{barreiro2021natural} introduced a particle-based viscoplasticity model with ultrasound haptic feedback, although the fidelity of finger-based interactions remained insufficient for detailed 3D modeling tasks. Additionally, physical proxies have been employed to enhance the virtual clay experience~\cite{sheng2006interface, pihuit2008hands, marner2010augmented}, such as combining direct finger input with deformable physical devices to enable clay-like sculpting operations, including deforming, smoothing, pasting, and extruding~\cite{sheng2006interface}. Efforts to adapt virtual clay modeling to VR environments have been limited. For instance, \citet{jang2014airsculpt} and \citet{moo2021virtual} made preliminary attempts to integrate these techniques into VR; however, their approach faced challenges in achieving the robustness and accuracy needed for intricate hand-object interactions.
To overcome these challenges, our work introduces an intuitive, contact-based VR modeling framework leveraging MPM \cite{jiang2016material} and 3D GS \cite{kerbl20233d}. Our approach significantly enhances real-time performance and hand-object interaction accuracy, enabling a more natural and immersive experience that underscores the unique benefits and potential of clay-like modeling in virtual environments.


\paragraph{Physics-Aware Interactions in VR}
Physics-aware hand-object interactions in VR have been extensively studied, enabling users to engage with various virtual phenomena such as particle-based animations~\cite{arora2019magicalhands}, deformable objects~\cite{moo2021virtual, deng2023phyvr, jiang2024vr}, and fluids~\cite{eroglu2018fluid}. These approaches often utilize hand-tracking data to map human hand motions to rigged virtual hand models, enabling dynamic interactions in immersive environments. For example, \citet{kumar2015mujoco, lougiakis2024comparing, holl2018efficient} use tracked hand information to define the pose of the virtual hand in each frame while incorporating frictional contact, achieving realistic hand-rigid-body interactions. Building on this, \citet{verschoor2018soft, jacobs2011soft, smith2020constraining} simulate soft hands using nonlinear soft tissue models, which allow for more natural and realistic interactions.
While these works primarily focus on interactions between hands and rigid objects, research on hand-deformable-object interactions has also gained traction. For instance, \citet{deng2023phyvr} developed PhyVR, a unified particle system for freehand interactions with multiple virtual materials; VR-GS~\cite{jiang2024vr} facilitates handle-based interaction with physically simulated elastic objects represented by 3D Gaussian Splatting~\cite{kerbl20233d}. However, these efforts largely emphasize physics-aware interactions rather than shape editing, which holds significant promise for creative tasks such as animation control~\cite{arora2019magicalhands}. To address this gap, we focus on physics-based hands-on 3D modeling in VR, leveraging both contact- and gesture-driven interactions to enable intuitive creation and editing of deformable objects for creative applications.

\paragraph{3D Modeling Tools in VR}
Creative tools for 3D modeling in VR encompass different categories. Among drawing-based modeling tools, Surface Drawing~\cite{schkolne2001surface} stands as a pioneering approach, enabling users to draw strokes in space using hand motions and edit 3D models with tangible tools. To enable smoother 3D curve creation, Dynamic Dragging~\cite{keefe2008tech} introduces an adaptive drag line that adjusts based on curvature and drawing speed. SurfaceBrush~\cite{rosales2019surfacebrush} and AdaptiBrush~\cite{rosales2021adaptibrush} convert dense collections of artist-drawn stroke ribbons into user-intended manifold free-form 3D surfaces. Cassie~\cite{yu2021cassie} is a conceptual modeling system in VR that leverages free-hand mid-air sketching and a novel 3D optimization framework to create a connected curve network. In addition, inspired by FiberMesh~\cite{nealen2007fibermesh}, RodMesh~\cite{schulz2019rodmesh} replaces the curve drawing with two-handed bending and stretching of virtual rods, allowing users to define outline shapes that are subsequently inflated into manifold mesh surfaces. \citet{peng2020autocomplete} presents a keyframe-based animated sculpting system that autocompletes user edits through an intuitive brushing interface. Commercial tools, such as Shapelab, Adobe Substance 3D Modeler\footnote{\url{https://www.adobe.com/products/substance3d/apps/modeler.html}}, and Kodon\footnote{\url{https://www.kodon.xyz/}}, mainly focus on geometric editing through polygonal or voxel-based sculpting. These tools allow users to manipulate geometry regions and apply operations with dynamic topology updates. Notably, Gravity Sketch introduces subdivision modeling, enabling users to create complex meshes from simple base topology while maintaining smoothness. Inspired by these prior works, we aim to further reduce the demand for professional 3D modeling expertise with physics-aware hands-on shape modeling, making 3D modeling in VR more accessible to novice users.





%% file: chapter03-design.tex
\section{Design Rationale}
To make hands-on 3D modeling in VR responsive, immersive, and user-friendly, we have derived the following design requirements and functionalities:
\begin{enumerate}
    \item \textbf{Realistic Physics-based Simulation.} This provides users with an intuitive understanding of how their input translates into deformations, offering a more intuitive shape modeling process compared to conventional desktop-based tools. 

    \item \textbf{Real-Time Performance.} Maintaining real-time frame rates and smooth rendering is crucial in VR to avoid motion sickness\footnote{VR motion sickness: a condition where users experience nausea and discomfort due to mismatches between visual and physical motion}, which severely impacts the user experience. Ensuring real-time responsiveness for shape deformations to hand input is essential for intuitive interactions.

    \item \textbf{Diverse Input Modalities.} In real-world clay modeling, people combine hand manipulation with tools like ball styluses and knives for precise and controlled deformations. Supporting the use of various tools in VR, beyond just hand manipulation, would enhance efficiency and unlock more creative shape-modeling possibilities.
    
    \item \textbf{Adhering to established operations.} Design is an iterative process, where complex models are often composed of multiple components. An effective modeling workflow should follow established operations, including creating, moving, scaling, merging, and copying individual objects, to achieve the final design.
    
    \item \textbf{Intuitive Spatial Interface.} A clear and user-friendly spatial interface is critical to guide users through the modeling process and improve the overall usability of the system.
\end{enumerate}
Simultaneously achieving all these goals presents a significant challenge. The more accurate the simulation and rendering, the greater the demand for computational resources. Appropriate input modalities are required to enable hands to deform virtual objects in mid-air as expected. In the following sections, we will explain how we built VR-Doh, including the overall system architecture, methodologies, and the technical trade-offs and innovations we employed.

%% file: chapter04-method.tex
\section{Method} \label{sec:method}

\begin{figure*}[!h]
    \centering
    \includegraphics[width=0.8\linewidth]{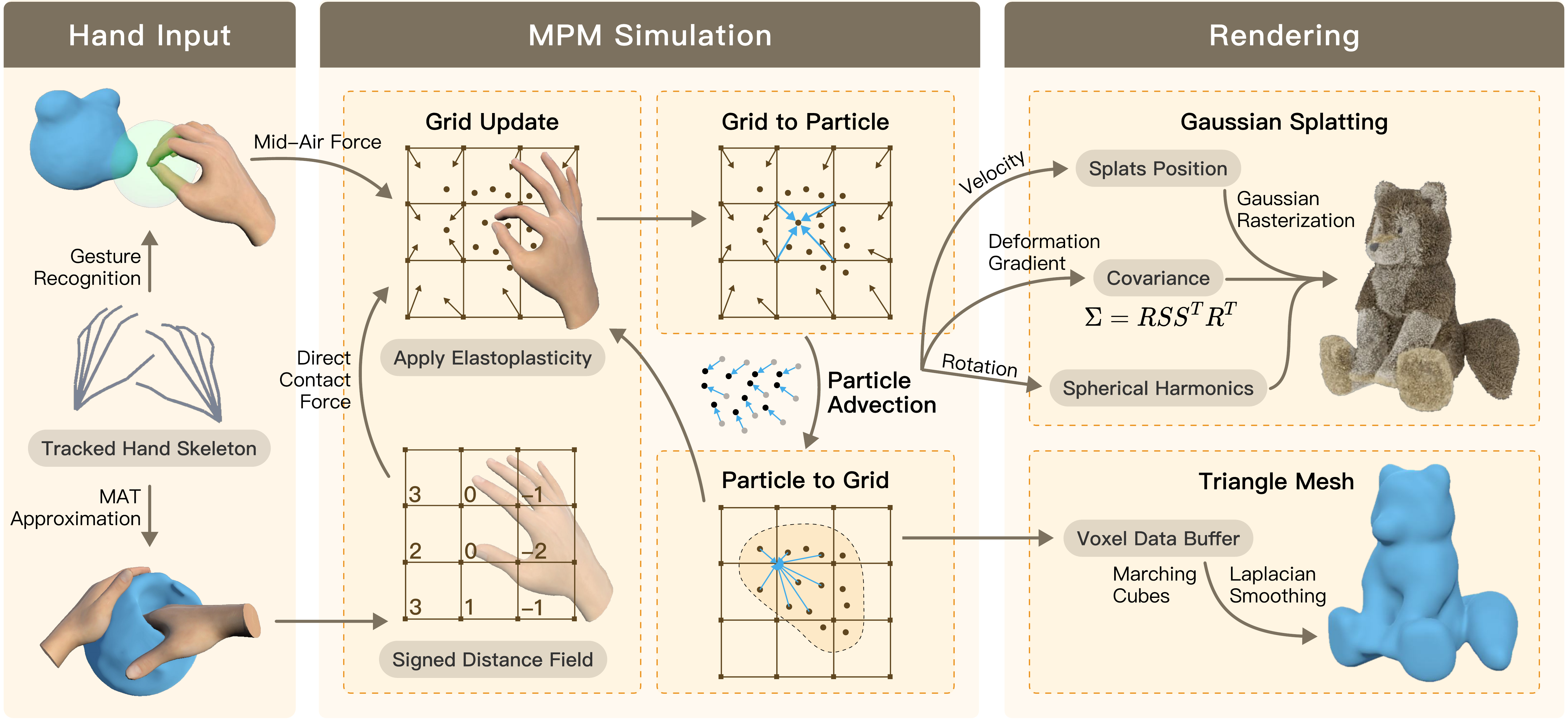}
    
    \caption{  \textbf{VR-Doh Pipeline.} Our interactive system enables hands-on 3D modeling in VR through contact- and gesture-based inputs. The pipeline integrates MPM to simulate realistic elastoplastic deformations and supports rendering using both 3D Gaussian Splatting and meshes, ensuring broad applicability across creative and practical modeling tasks.}
    \label{fig:system}
\end{figure*}

Our system, built on PhysGaussian~\cite{xie2024physgaussian}, integrates intuitive user interactions, real-time simulation, and high-quality rendering to enable hands-on 3D modeling in VR. PhysGaussian combines the Moving Least Squares (MLS) Material Point Method (MPM)~\cite{hu2018moving} and 3D Gaussian Splatting~\cite{kerbl20233d} for elastoplastic simulation and photorealistic rendering. The technical background of these foundational methods is provided in our supplemental document. However, directly applying these techniques does not fully address the challenges of enabling real-time performance, intuitive usability, and dynamic updates in our system. To overcome these limitations, we introduce key innovations, including particle-level collision handling and localized simulations for efficient and accurate modeling (\autoref{sec:simulation}), as well as decoupled appearance and physical representations and uniform Gaussian volume regularization for enhanced rendering (\autoref{sec:rendering}). These innovations, combined with natural interaction modes such as contact- and gesture-based inputs that mimic real-world actions (\autoref{sec:input}), make VR-Doh both accessible to novice users and powerful for detailed modeling tasks. \autoref{fig:system} and Algorithm~\ref{alg:pipeline} provide an overview of the system framework.

\begin{algorithm}\small
\caption{VR-Doh Pipeline}\label{alg:pipeline}
\begin{algorithmic}[1]
\While{True}
    \State Update\_Input() \Comment{\autoref{sec:input}}
    \For{each substep}
        \State Sim\_substep()\Comment{Supp. Doc. and \autoref{sec:simulation}}
    \EndFor
    \State Update\_Rendering()\Comment{\autoref{sec:rendering}}
\EndWhile
\end{algorithmic}
\end{algorithm}


\subsection{Hand-based User Interaction}\label{sec:input} 



\begin{figure}[htbp]
    \centering
    \includegraphics[width=\linewidth]{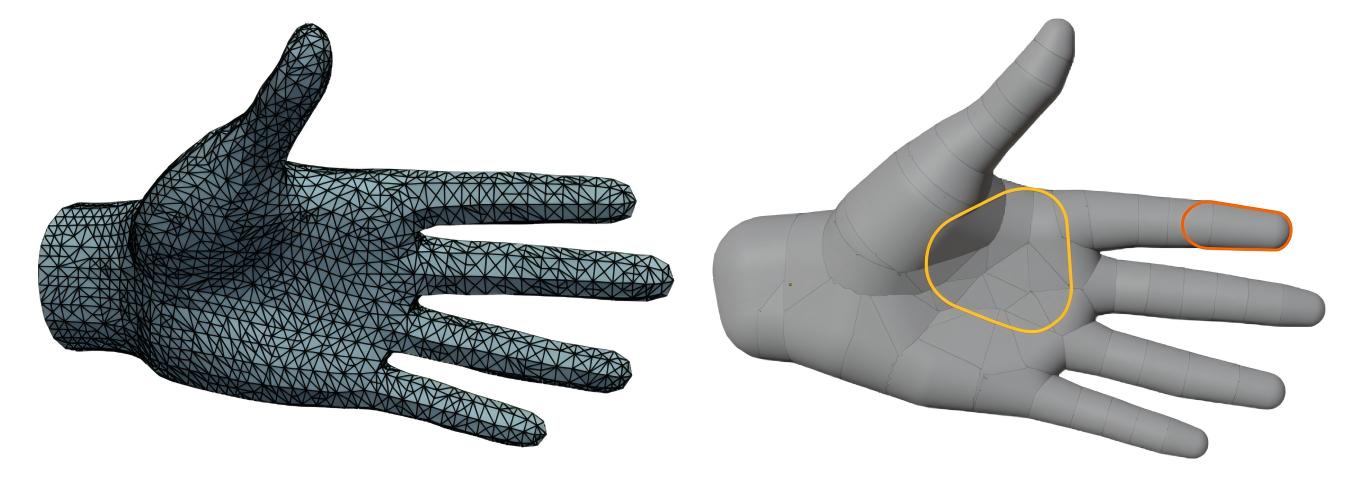}
    \captionof{figure}{  \textbf{Medial axis approximation of hand geometry.} The highlighted curves indicate the boundaries of a specific medial slab and a medial cone.}
    \label{fig:mat_hand}
    \label{fig:Improved_gaussian}
\end{figure}

\paragraph{Contact-based Modeling}
To enable efficient contact handling for interactive feedback, we approximate the geometry of hands and integrated deformation tools using the medial axis transform (MAT) \cite{faraj2013progressive, stolpner2011medial}. The medial axis of a 3D model consists of points that are centers of maximally inscribed spheres. By incorporating radius information along the medial axis, the MAT can reconstruct the original geometry. MAT can be discretized as a small number of linearly interpolated spheres or medial primitives while precisely describing the shape. These medial primitives are classified as either medial cones or slabs~\cite{sun2015medial}, defined by two or three vertices on the medial axis. In our implementation, we use the quadratic error metric \cite{li2015q} to compute the medial mesh of hands and tools. Specifically, a hand can be approximated by 76 medial cones and 28 medial slabs (Figure~\ref{fig:mat_hand}). For collision handling during simulation, we adopt the method from Medial IPC~\cite{lan2021medial} to calculate distances between MPM grid nodes and medial primitives representing hands or tools. Additionally, we employ the bounding boxes of medial primitives to construct a spatial hash data structure, enabling efficient collision detection.

\begin{figure}[htbp]
    \centering
    \includegraphics[width=\linewidth]{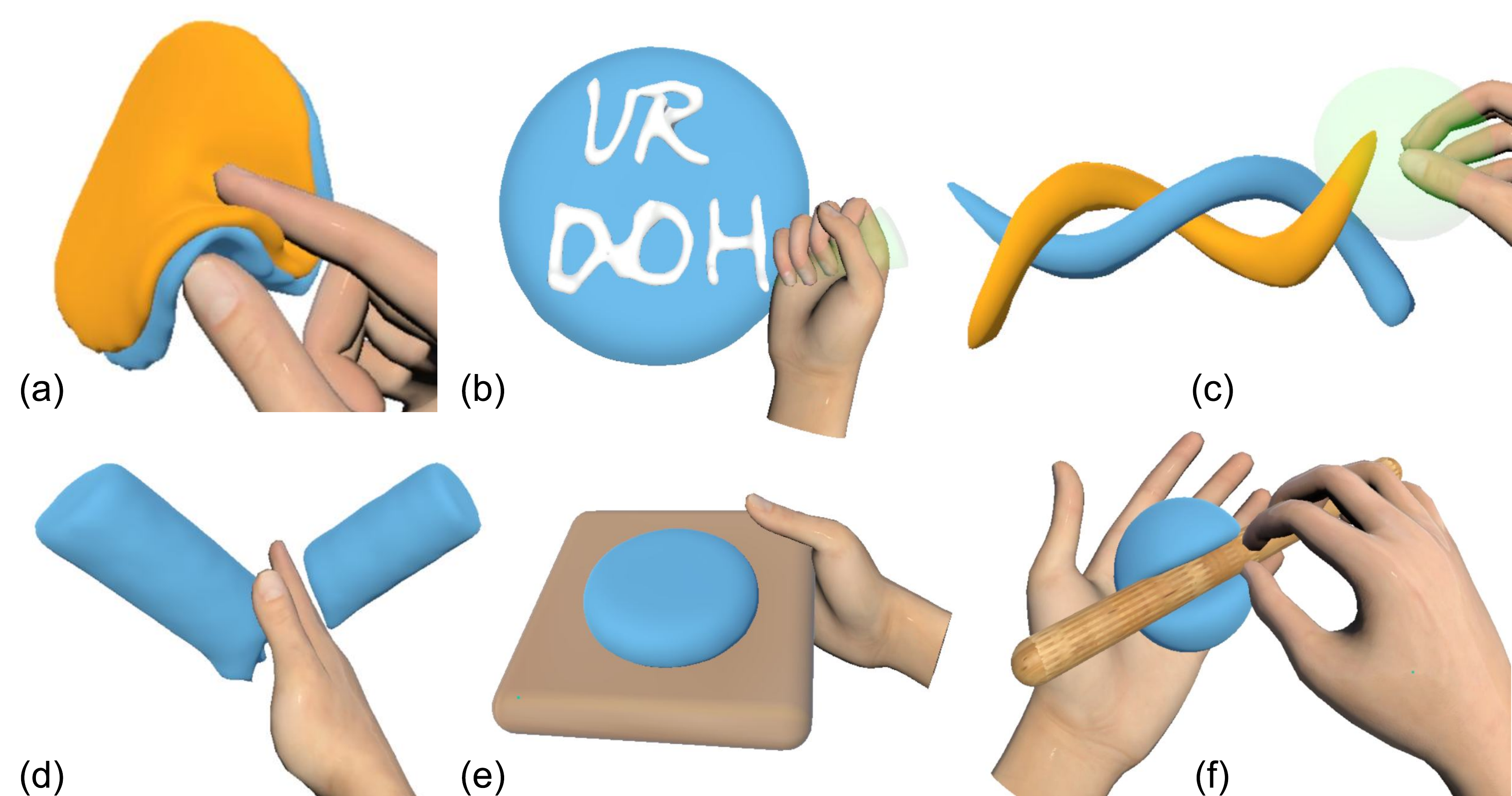}
    
    \captionof{figure}{  \textbf{Featured Operations:} (a) Pinching and reshaping with fingers; (b) Using a sourcing tool to extrude new materials onto an object; (c) Mid-air gestures for bending and twisting objects; (d) Hand-based cutting of objects; (e) (f) Tool-based operations with a slab and a rod; .}
    \label{fig:Featured_Operations}
\end{figure}

Beyond direct hand manipulation, we provide auxiliary modeling tools such as a planar slab, a rod, and scissors (Figure~\ref{fig:Featured_Operations}) to support diverse and controlled deformations, mimicking creative processes observed in the real world. Upon selection, each tool attaches to a hand joint, allowing users to control it through hand movements. Each hand operates independently, enabling seamless transitions between different tools. During hand tracking, the position and orientation of each medial primitive are determined by the nearest joints of the hand skeleton. To enhance tracking stability, we apply a moving average kernel to the data from the past 0.2 seconds, with weights determined by frame duration. This ensures consistent and reliable hand-tracking performance across varying frame rates.

Overall, our efficient contact handling techniques enable interactive modeling operations on deformable objects, such as pinching, squeezing, and folding, while avoiding noticeable visual artifacts. We use the hand texture from \citet{Pohl2022Hands} for rendering.


\paragraph{Mid-air Gestural Modeling}
Mid-air gestural input is advantageous when users face difficulties selecting the desired editing area on an object using contact-based modeling, which helps to avoid unintended changes to the object. We provide a mid-air pinch gesture (where the tip of the thumb touches the tip of the middle finger), enabling users to perform operations such as stretching or twisting materials by applying a force field to selected MPM particles. A green highlight is used to visualize the selection area and the applied force field, providing intuitive visual feedback. Users can further refine the radius of the selection area and adjust the force magnitude for more precise control, as shown in \autoref{fig:Featured_Operations}.

\paragraph{Other Operations}
Our user interface provides a range of common operations to enhance usability and ensure smooth modeling workflows. Objects in the scene can be \textit{selected} and \textit{moved} by grabbing their bounding boxes with the hands or by using a ray-casting gesture to intersect the bounding box. \textit{Scaling} operations are performed by grasping the object’s bounding box with both hands and moving them outward or inward, visually scaling the object up or down without altering its physical size. This zooming functionality enables seamless transitions between fine-grained local editing and broader global shaping.
To create new objects, users can \textit{load} built-in primitive shapes, such as spheres, tori, or cubes, or utilize a \textit{sourcing} tool that extrudes geometry with customizable cross section (e.g., circles, squares, or stars). The extrusion process dynamically samples MPM particles, with the direction and speed controlled by hand movements.
Additionally, the system integrates commonly used operations in 3D modeling tools, such as \textit{Merge}, \textit{Copy}, \textit{Reset}, and \textit{Delete}, allowing users to efficiently manipulate modeled objects.

\subsection{Simulation}\label{sec:simulation}

\begin{algorithm}[htb]\small
\caption{Sim\_substep() \ \ \textit{(More details in our supp. doc.)}}
\label{alg:mpm}
\begin{algorithmic}[1]
\State Particle\_to\_Grid()
\State Update\_Grid\_Velocity()
\State Grid\_to\_Particle()
\State Particle\_Projection() \Comment{\autoref{sec:simulation}}
\State Apply\_Plasticity()
\end{algorithmic}
\label{alg:mpm}
\end{algorithm}

\paragraph{Particle-Level Collision Handling}
Due to the limited resolution of the simulation grid, MPM particles can penetrate the boundaries of passive objects, such as hands and tools, despite applying grid-based boundary conditions. Increasing grid resolution to fully resolve these geometries is computationally expensive and impractical. To address this issue, we introduce a particle-level collision handling step that operates independently of grid resolution.

For particles that remain inside passive objects after their states are updated using grid information in the \texttt{\small Grid\_To\_Particle} step, we project them out and adjust their velocities:
\begin{equation}
\begin{aligned}
\mathbf{x}_p = \mathbf{x}_p + \mathbf{p}_p, ~~~~\text{and}~~~~ \mathbf{v}_p = \mathbf{v}_{\text{boundary}}, 
\end{aligned}
\end{equation}
where $\mathbf{p}_p$ is the vector connecting the particle position $\mathbf{x}_p$ to the closest point on the boundary, and $\mathbf{v}_{\text{boundary}}$ is the velocity of this closest point. This step ensures particles remain outside the passive objects and conform to their surface geometry, enabling accurate hand-object interactions even at moderate grid resolutions.

\paragraph{Localized Simulation}
While MPM is effective for large deformations, real-time simulation is constrained by typical hardware capabilities, supporting approximately 100K particles. However, large-scale Gaussian splatting scenes often contain over 500K particles, making full-scene simulation infeasible in real time. To address this, we propose a localized simulation approach tailored to user interactions. Since the primary input comes from hand tracking, our system confines the simulation to a cubic region centered around the user’s hands, leaving particles outside this region stationary. This approach significantly reduces computational costs, enabling higher frame rates without 
compromising user experience. 

\subsection{Rendering}\label{sec:rendering}


\paragraph{Uniform Gaussian Volume Regularization}
The original 3D Gaussian Splatting (GS) method captures fine details only on the surface of objects, while the interior is typically empty or filled with a small number of large Gaussians. While this approach is sufficient for static objects, large deformations can expose the interior, leading to blurry rendering artifacts. 
PhysGaussian~\cite{xie2024physgaussian} attempts to address this by sampling additional Gaussians inside the object and assigning them the color of the closest surface Gaussian. However, this strategy can still result in blurry visuals in the interior during significant deformations.

To tackle this issue, we introduce a loss function during the training of 3D GS to penalize large volume differences of Gaussians:
\begin{equation}
L_{\text{vol\_ratio}} = \max \left( \frac{\text{mean}(V_{\text{top,} \alpha})}{\text{mean}(V_{\text{bottom,} \alpha})}, r \right) - r,
\end{equation}
where $V$ represents the average volume of the top $\alpha$ percent of Gaussians with the largest or smallest volumes, and $r$ is the target maximum allowable volume ratio. In practice, we find that setting $r = 2$ and $\alpha = 30\%$ yields effective results. This loss encourages Gaussians to have more uniform volumes, reducing the likelihood of blurry artifacts and ensuring consistent visual fidelity even under large deformations.

\paragraph{Decoupled Appearance and Physical Representations}
In \citet{xie2024physgaussian}, each Gaussian is treated as an MPM particle. To accurately render complex appearances, the density of Gaussians is typically much higher than the precision required by MPM for simulating elastoplastic deformations in the context of 3D modeling. To address this mismatch, we decouple MPM particles for simulation from Gaussian kernels used for rendering. A smaller number of MPM particles drive a larger number of Gaussians through the MPM grid, maintaining rendering accuracy while significantly improving simulation efficiency.
Specifically, during each simulation substep, MPM particles execute the full \texttt{\small Sim\_Substep} first. Subsequently, Gaussian kernels use the grid velocity information from the MPM simulation to perform \texttt{\small Grid\_to\_Particle}, \texttt{\small Particle\_Projection}, and \texttt{\small Apply\_Plasticity}. These steps are fully parallelizable and account for only a small portion of the total computational cost in a typical MPM simulation. This decoupling strategy optimizes computational efficiency without compromising visual fidelity, ensuring a seamless balance between simulation and rendering.

\paragraph{Mesh Rendering}
In addition to supporting the editing of 3D objects represented by Gaussian splatting, our system allows users to create models from scratch by reconstructing surface meshes from MPM particles using the Marching Cubes algorithm. Specifically, we compute the density field of the modeled objects by transferring particle mass onto a uniform grid during the \texttt{\small Particle\_to\_Grid} step and extract an isosurface to construct the mesh. This grid is independent of the simulation grid, following the decoupled appearance and physical representations. To improve surface quality, Laplacian smoothing is applied to the resulting mesh.
%
The original Marching Cubes algorithm may create unwanted connections or "sticking" between regions of different objects that are in close proximity. To address this, a "category" attribute is assigned to each particle. During the \texttt{\small Particle\_to\_Grid} transfer, separate density fields are maintained for each particle category. The Marching Cubes algorithm is then applied independently to each density field, ensuring that each category is reconstructed as an independent mesh and rendered separately.

\subsection{System Implementation}

We developed our system using Unity, integrating 3D Gaussian Splatting and Marching Cubes techniques. For the simulation component, we utilized Taichi \cite{hu2019taichi} as the compilation engine, which is well-suited for developing high-performance physics-based simulators. The simulation code was compiled into binary files and imported into Unity, where it was executed via C\# bindings.
For experiments, the system ran on a 16-core 4.3GHz AMD Ryzen 9 9950X machine with an Nvidia RTX 4090 GPU, paired with a Meta Quest 3 VR headset featuring hand-tracking functionality as the display and user input device. This setup enabled efficient and responsive interactions within the VR environment.

%% file: chapter05-evaluation.tex
\section{Evaluation}
In this section, we evaluate the real-time performance, visual consistency, and realism of our system, highlighting the contributions of our technical innovations with ablation studies. Unless stated otherwise, the experimental setup is as follows: The simulation uses a grid resolution of \(64^3\) with 8 particles per grid cell at initialization. Each frame performs 5 simulation substeps, employing a neo-Hookean elasticity model (Young's modulus \(E=10^6 Pa\), Poisson's ratio \(\nu=0.3\))  combined with a von Mises plasticity model (yield stress \(\sigma_y=10^5 Pa\)). For rendering, the density field is sampled at \(128^3\), with mesh reconstruction via the Marching Cubes algorithm and 5 iterations of Laplacian smoothing to improve surface quality. To ensure consistent comparisons, we recorded a specific hand trajectory for each example and applied it across all tests with varying parameters.

\begin{figure*}[ht]
    \centering
        \includegraphics[width=0.8\linewidth]{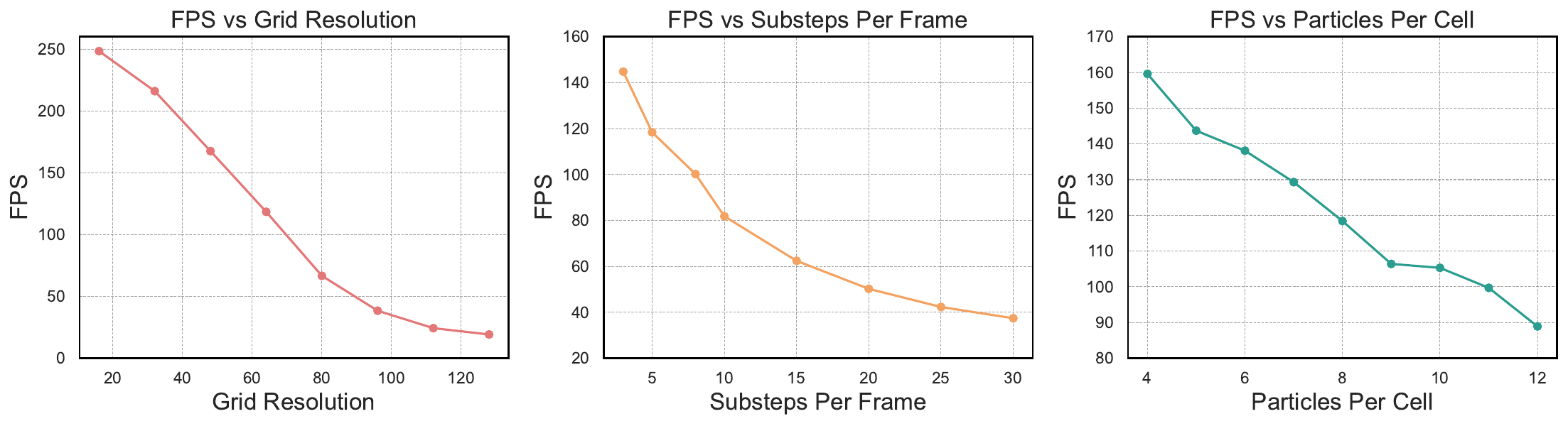}
        
    \captionof{figure}{\textbf{Real-Time Performance Evaluation.} FPS of VR-Doh with varying simulation grid resolution, substeps per frame, and MPM particles per cell in a sphere compression scenario (\autoref{fig:Boundary_Projection}, bottom). Results demonstrate VR-Doh's ability to maintain real-time performance under high computational demands.}
    \label{fig:performance}
\end{figure*}

\begin{figure}[ht]
    \centering
    \includegraphics[width=0.9\linewidth]{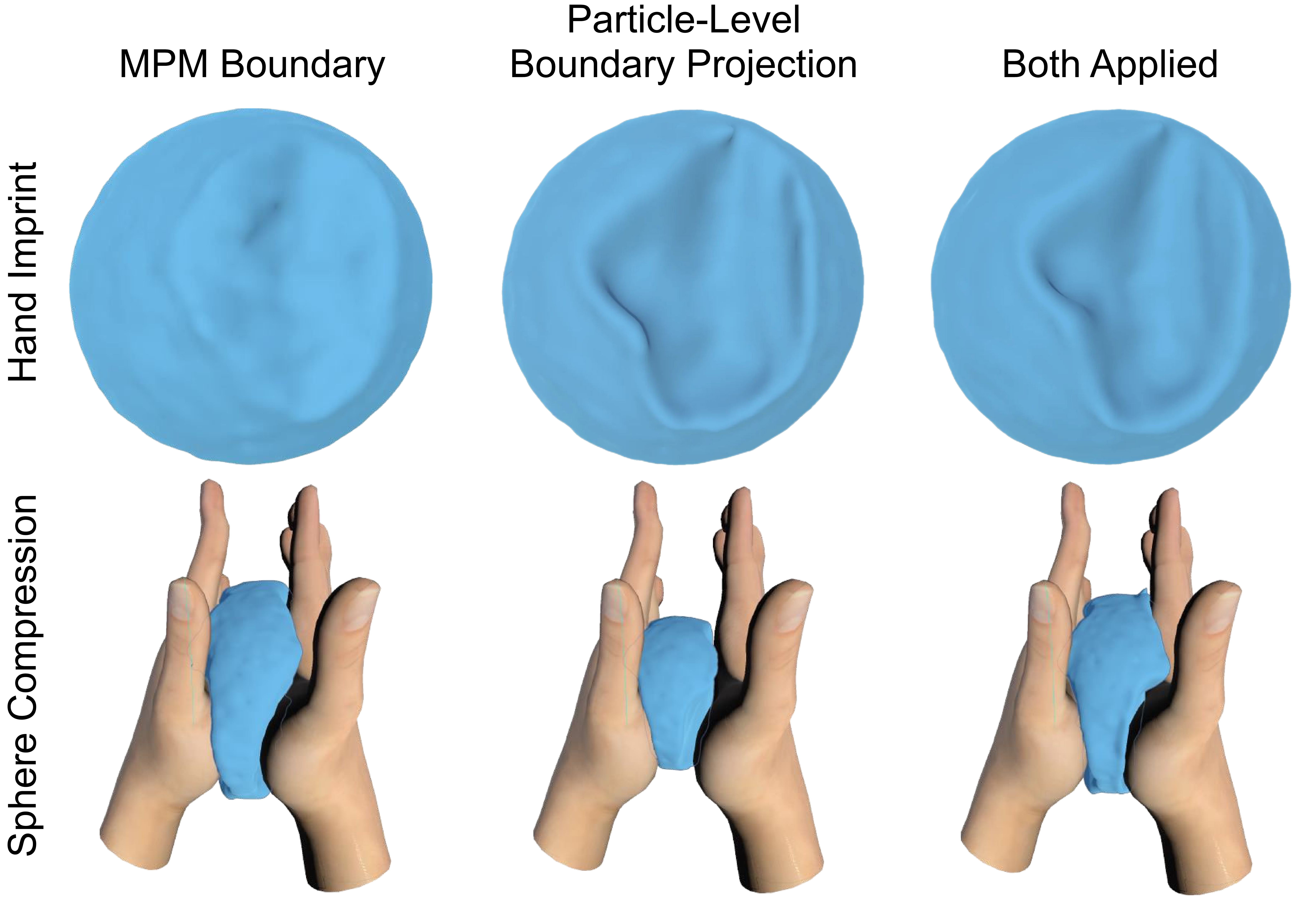}
        
    \captionof{figure}{ {\textbf{Particle-Level Collision Handling.} Comparison of MPM boundary conditions with and without particle-level collision handling on a hand imprint (top) and a sphere compression (bottom) example. Combining both methods (right) achieves realistic deformations with enhanced detail while avoiding volume loss.}}
    \label{fig:Boundary_Projection}
\end{figure}

\paragraph{Real-Time Performance}
To evaluate the real-time performance of VR-Doh, we conducted an experiment involving a simple scenario where a sphere is compressed between two hands (as shown in \autoref{fig:Boundary_Projection}, bottom row). To analyze performance scalability, we varied the simulation grid resolution, the number of substeps per frame, and the number of particles per grid cell, while keeping other parameters constant. The overall FPS under these conditions is summarized in \autoref{fig:performance}. The results demonstrate that VR-Doh can maintain real-time performance, achieving interactive frame rates even with higher computational demands, such as a \(100^3\) simulation grid resolution, 30 substeps per frame, and 12 particles per cell.

\begin{figure*}[htbp]
    \centering
        \includegraphics[width=0.65\linewidth]{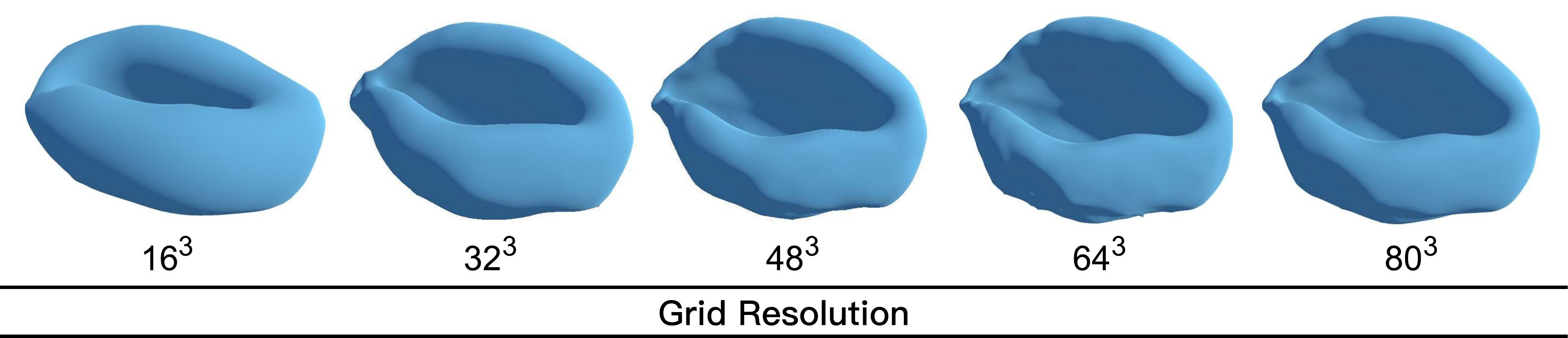}
        
    \captionof{figure}{\textbf{Convergence under Spatial Refinement.} Sphere deformations under identical hand imprints and squeezing converge at resolutions above $48^3$, ensuring accuracy without excessive computational cost.}
    \label{fig:reso}
\end{figure*}

\paragraph{Convergence Under Spatial Refinement}
To validate the consistency of our simulation, we conducted an experiment where a hand trajectory presses into a sphere to create a hole and then squeezes it from both sides. The simulation was tested with varying grid resolutions while keeping the number of particles per cell constant. As shown in \autoref{fig:reso}, the deformations converge once the grid resolution exceeds \(48^3\), indicating that our simulation achieves consistent and accurate results without requiring excessively high resolutions.

\paragraph{Particle-Level Collision Handling}
In \autoref{fig:Boundary_Projection}, we compare the traditional MPM boundary condition with and without our particle-level collision handling using a hand imprint (top row) and a sphere compression (bottom row) example. As shown in the figure, applying particle-level collision handling produces significantly clearer imprints, enhancing the ability to adjust finer details. However, the results in the second row demonstrate that using only particle-level collision handling can lead to severe volume loss, as the particle projection step does not account for elastoplasticity forces. By combining both methods, our approach achieves realistic overall deformations while enabling precise adjustments for detailed modeling.

\paragraph{Localized Simulation}
\begin{table}[H]
\centering
\caption{ \textbf{Localized Simulation.} Comparison of FPS with and without localized simulation in large 3D Gaussian Splatting scenes, showing an average frame rate improvement of $\mathbf{3.3\times}$ with localized simulation enabled.}
\label{tab:local}

\begin{tabular}{>{\centering\arraybackslash}m{1cm}|>{\centering\arraybackslash}m{1.5cm}|>{\centering\arraybackslash}m{2.4cm}|>{\centering\arraybackslash}m{2.2cm}}
\toprule
\textbf{Scene} & \textbf{Splats count} & \textbf{FPS w/o localized  sim.} & \textbf{FPS w localized sim.} \\
\hline
Garden & 660K & 12.2 & \textbf{39.8} \\
\hline
Bicycle & 480K & 13.1 & \textbf{43.5} \\
\hline
Room & 382K & 13.6 & \textbf{44.5} \\
\bottomrule
\end{tabular}
\end{table}

We evaluated VR-Doh's performance with and without localized simulation across several typical 3D GS scenes, including Garden, Bicycle, and Room, each containing several hundred thousand Gaussians. During the tests, users navigated through each scene and interacted with primary objects using hand gestures, introducing deformations. As shown in \autoref{tab:local}, enabling localized simulation resulted in an average frame rate improvement of $3.3\times$, demonstrating its significant impact on performance.

\begin{figure}[ht]
    \centering
        \includegraphics[width=\linewidth]{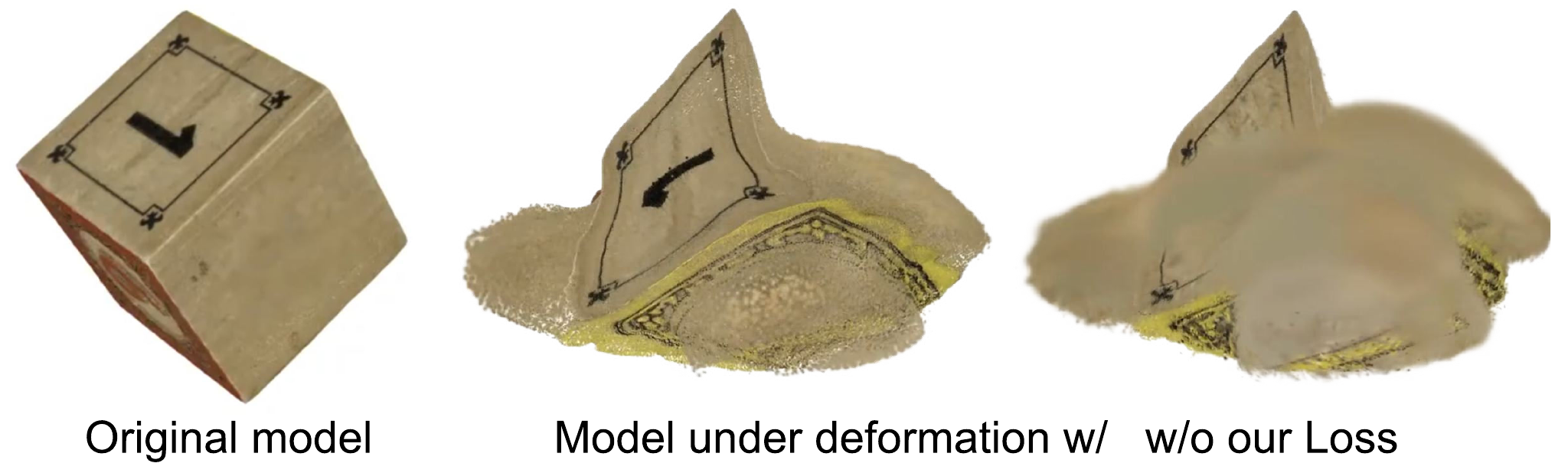}
        
    \captionof{figure}{  \textbf{Uniform Gaussian Volume Regularizer.} Comparison of a sand cube dropped under gravity shows that applying the regularization (middle) preserves sharp details and avoids blurry artifacts in the interior (right), ensuring high visual fidelity during large deformations.}
    \label{fig:Regularizer}
\end{figure}

\begin{figure}[htbp]
    \centering
    \includegraphics[width=1.0\linewidth]{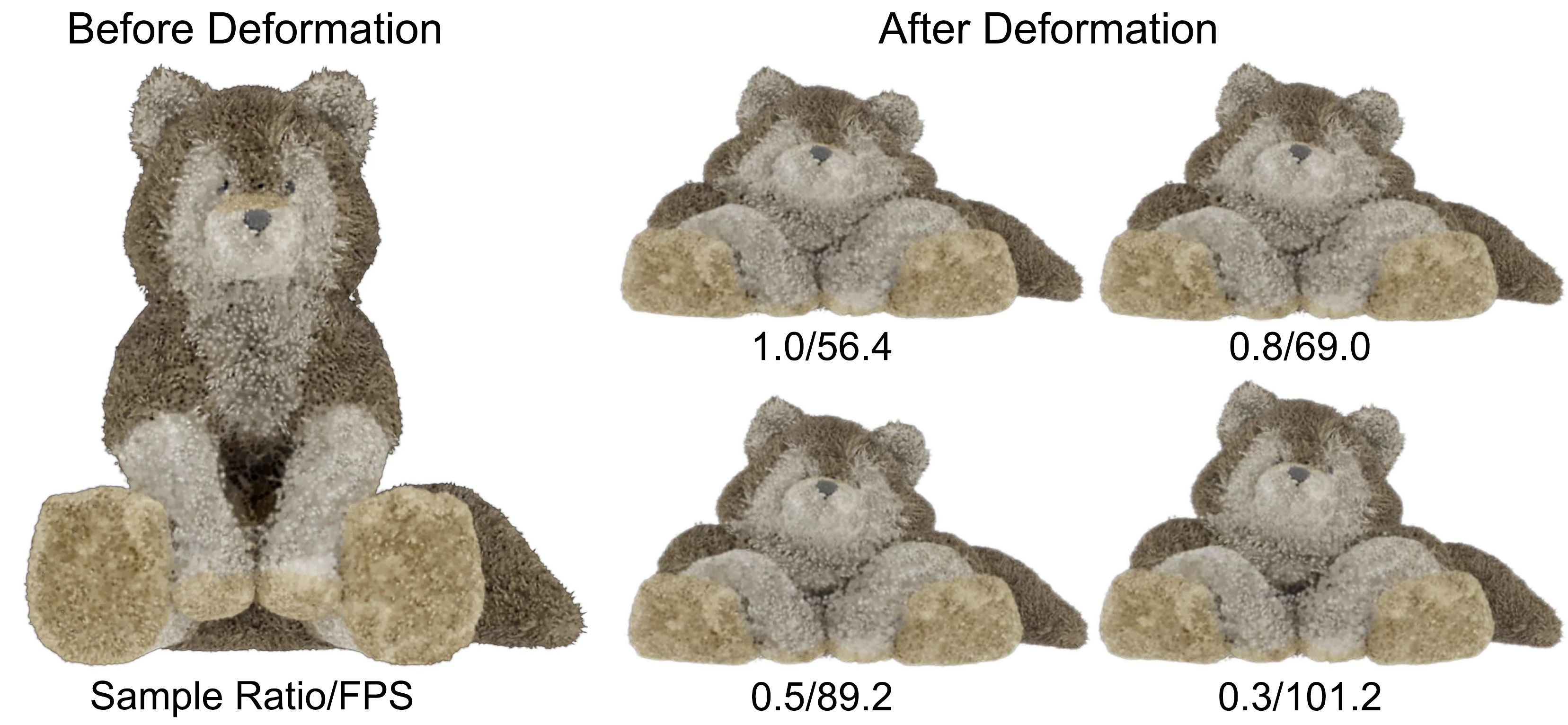}
    \caption{ {\textbf{Decoupled Appearance and Physical Representations.} This example of a wolf sitting under gravity demonstrates that reducing the number of MPM particles driving the 3D Gaussians slightly increases the wolf's volume but maintains high visual fidelity while significantly improving FPS, showcasing the effectiveness of our approach. }}
    \label{fig:Improved_gaussian}
\end{figure}

\paragraph{Uniform Gaussian Volume Regularizer}
We evaluated the effectiveness of our Uniform Gaussian Volume Regularization by simulating a textured sand cube represented with 3D Gaussians as it was dropped onto the ground under gravity, resulting in significant deformations. \autoref{fig:Regularizer} compares the original model (left) with the deformed results using (middle) and without (right) the proposed volume regularization loss, both trained for 30,000 iterations using Adam \cite{Kingma2014AdamAM}. Without the loss, the exposed interior reveals noticeable blurry artifacts caused by uneven Gaussian volumes. In contrast, applying the regularization maintains consistent Gaussian volumes, preserving sharp details and achieving high visual fidelity even under extreme deformation. 
\paragraph{Decoupled Appearance and Physical Representations}
We evaluated the performance of our method under different sampling ratios between MPM particles and Gaussian kernels in a simulation where a wolf sits on the ground under gravity. Here, we maintain the MPM particle per cell, so the simulation grid resolution decreases as we downsample MPM particles. \autoref{fig:Improved_gaussian} illustrates the original (left) and deformed (right) configurations and the FPS achieved with varying sampling ratios. The results show that as the sampling ratio decreases—fewer MPM particles driving the 3D Gaussians—the wolf's volume slightly increases while maintaining high visual fidelity and preserving detailed features. This behavior is expected, as lower resolutions in MPM typically exhibit stiffer deformation due to discretization errors. Notably, by reducing MPM particles to approximately $30\%$ of the Gaussian kernels, the FPS nearly doubles, demonstrating the effectiveness of our approach to decouple appearance and physical representations, achieving significant performance gains without sacrificing visual quality.

%% file: chapter06-case.tex
\section{Case Study}

\paragraph{Featured Operations}
Figure~\ref{fig:Featured_Operations} highlights featured modeling operations in VR-Doh, including contact-based shape editing with hands and tools like slabs and rods, mid-air gesture-based bending and twisting, and the sourcing tool for adding new materials. The objects exhibit realistic elastoplastic deformations, closely resembling real-world behavior. This physical realism enables users to intuitively model objects by drawing on their real-life experiences. Users can freely adjust material parameters such as stiffness and yield stress, or choose from a set of predefined materials to suit different modeling needs.

\begin{figure}[htbp]
    \centering
    \includegraphics[width=1.0\linewidth]{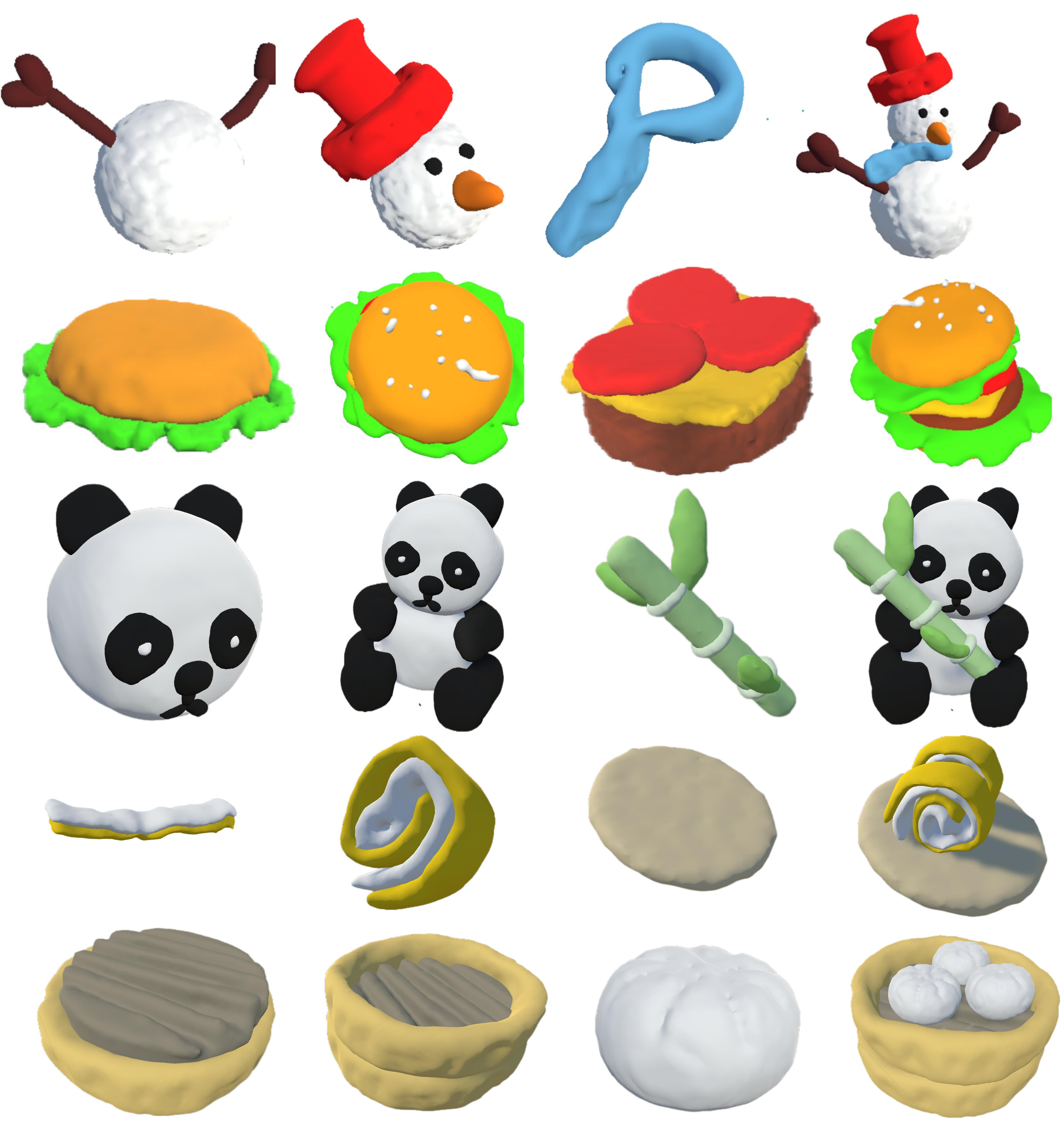}
    \caption{ {\textbf{Object Creation Examples.} Crafting a snowman, assembling a hamburger with gravity-stacked layers, creating a panda holding bamboo, rolling a Swiss roll, and shaping steamed buns in a bamboo steamer, showcasing intuitive and precise 3D modeling.
    }}
    \label{fig:mc_examples}
\end{figure}
\begin{figure*}[htbp]
    \centering
        \includegraphics[width=0.8\linewidth]{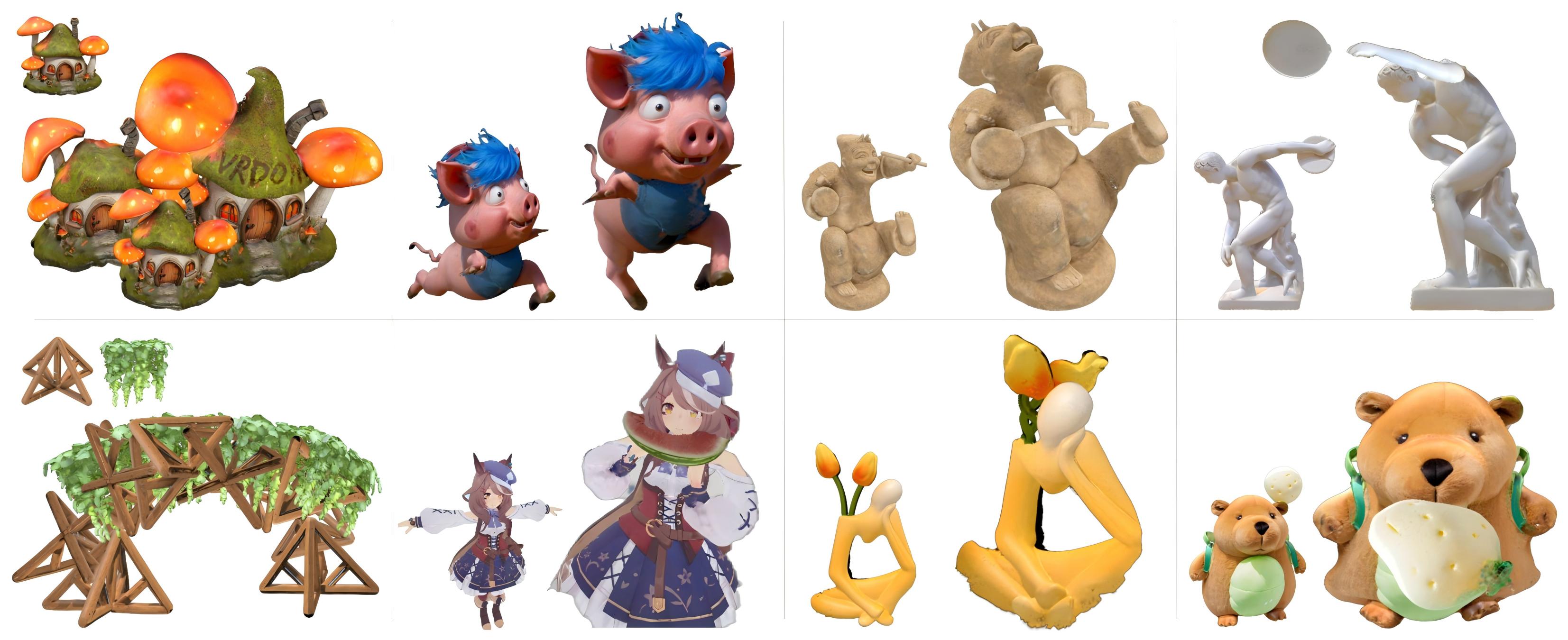}
        
    \captionof{figure}{\textbf{3D GS object editing examples.} Users combined, reshaped, and customized models using hand gestures and tools, bringing objects to life, merging elements seamlessly, and refining intricate details with ease. These examples highlight VR-Doh’s ability to simplify complex modeling tasks while preserving creativity and control. 
    }
    \label{fig:object_editing}
\end{figure*}

\paragraph{Object Creation}
We demonstrate object creation using VR-Doh, as shown in \autoref{fig:mc_examples}. 
In the first row, a snowman is assembled, starting with a blank snowfield, molding the body, and adding details. 
The second row demonstrates the creation of a hamburger by stacking individual components to form a layered structure.
In the third row, a panda is carefully modeled from head to body, incorporating bamboo as an accessory that seamlessly integrates with its hands.
The fourth row depicts the crafting of a Swiss roll, where layers are rolled together to replicate a realistic pattern.
Finally, the fifth row shows the formation of steamed buns, shaped and arranged within a bamboo steamer.
These examples highlight VR-Doh’s strength in simplifying tasks that require precise spatial arrangements, which are significantly more challenging to achieve using a mouse on a 2D screen. With VR-Doh, object creation becomes as natural and intuitive as manipulating items in the real world.
\paragraph{3D GS Object Editing}
\autoref{fig:object_editing} demonstrates 8 examples of editing GS-represented objects, showcasing the intuitive and creative capabilities of VR-Doh. 
First, multiple mushroom-shaped houses were assembled, with their roofs sharpened and the mushrooms enlarged through hand manipulation and mid-air pinch gestures. The text "VR-Doh" was then sculpted onto the roof by hand. 
Next, a red pig transitioned from running to performing ballet through flexible selection and dragging of specific body parts, with its head rotated to face the audience using the pinch gesture. 
Two statues were brought to "life": a terracotta figurine joyfully lifted its head and began drumming, while the Discobolus threw its discus with dynamic force. This involved using hands to cut and separate the discus before repositioning it. 
Subsequently, wooden frames were interlocked and freely stacked into an arch through gravity and contact interactions, then merged seamlessly with greenery. The greenery’s complex shapes were also easily refined by hand. 
A girl’s posture was adjusted as her arms bent to mimic the act of eating a watermelon. 
Following that, flowers on a vase were flexibly selected using the pinch gesture, then intertwined in space with penetrations automatically avoided, demonstrating precise and convenient control. 
Finally, a capybara plush toy was made to appear fatter, while its head decoration was reshaped into a strawberry and placed in the front. These examples highlight VR-Doh’s ability to simplify complex modeling tasks while maintaining creativity and controllability. 

%% file: chapter07-user.tex
\section{User Study}
To evaluate VR-Doh's usability, effectiveness, and potential applications, we conducted two user studies involving both novice and expert participants. The first study (\textit{US1}) focused on open-ended creative tasks to assess the system's intuitiveness, immersion, and efficiency in 3D modeling, and to gather participants' feedback and suggestions. The second study (\textit{US2}) directly compared VR-Doh with the conventional modeling software Blender, highlighting their respective strengths and limitations.

\subsection{US1: Usability and Effectiveness of VR-Doh}
To understand how experts and novices use VR-Doh for creative tasks in VR, we invited 6 participants (P1-P6) with 3-5 years of prior 3D modeling experience, and 6 participants without any prior experience (P7-P12).

\paragraph{Tasks \& Procedure}
Each participant was required to conduct two tasks, with a 5-minute break in between: 1) \textit{Editing a 3D Model}: Participants were free to modify a pre-existing 3D model represented by GS, e.g., reshaping a plush toy or a cartoon character, to match their design intention; 2) \textit{Creating a 3D Model from Scratch}: Participants can freely use primitive shapes such as spheres, cubes, tori, and cylinders to create a new 3D model, e.g., a snowman or a hamburger. Before starting, participants practiced basic operations with a provided shape until they became familiar with the system. After completing the tasks, participants filled out the USE questionnaire~\cite{lund2001measuring} to evaluate usability and completed the Simulator Sickness Questionnaire (SSQ)~\cite{kennedy1993simulator} to measure any discomfort experienced during the study. Finally, participants took part in a 10-minute semi-structured interview to provide more in-depth insights on usability.

\begin{figure*}[htbp]
        \includegraphics[width=0.8\linewidth]{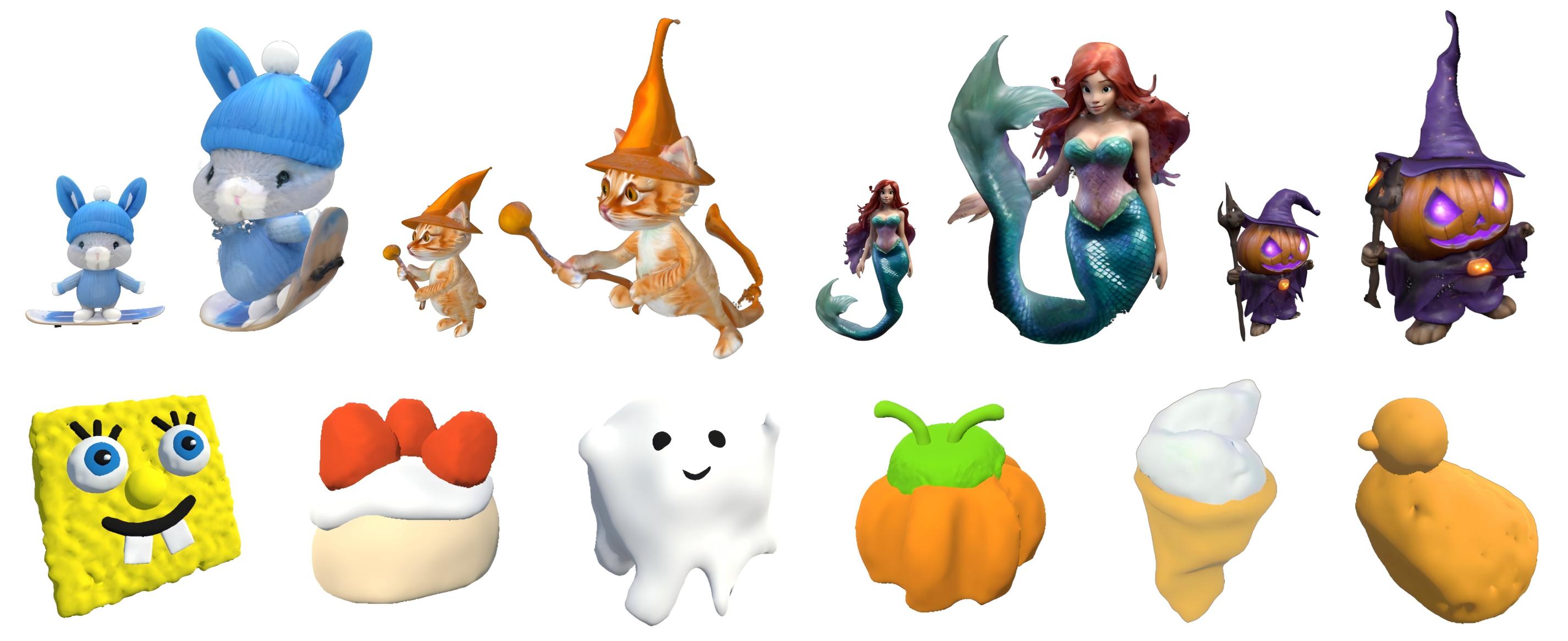}
        
    \captionof{figure}{\textbf{User Study Outcomes.} Edited 3D GS models (top row) and newly created models (bottom row) by novice and expert participants.}
    \label{fig:user_study_results}
\end{figure*}

\paragraph{Results}
Both expert and novice participants found VR-Doh to be highly intuitive, immersive, and efficient in their 3D modeling process, and they also claimed that the created shapes closely matched their initial design expectations. Some of their design outcomes are shown in \autoref{fig:user_study_results}. Participants rated the system highly on usefulness (5.6/7), ease of use (5.5/7), ease of learning (6.0/7), and overall satisfaction (5.9/7). Moreover, no discomfort and nausea were found. We also report their subjective feedback below.

\paragraph{Intuition and Realism in 3D Modeling} Our participants spoke highly of the low learning curve of VR-Doh, allowing them to quickly engage in shape editing or create 3D models with their hands. P1 and P3 mentioned that directly manipulating the shape of objects with their hand was more flexible than deforming the object by setting control points. P5 found it more convenient to modify the shape without the need for rigging, which is tedious in conventional software. P1 stated that the rod tool was particularly useful and entertaining for poking holes and shaping details on a strawberry shape, and said it felt like \textit{"walking inside the strawberry."} The mid-air pinch interaction was highlighted, which allows users to select the area of the object they want to deform or sculpt. P2 considered that the ability to freely rotate the viewpoint in VR is more intuitive than the traditional three-view approach.

In addition, our participants highly praised the modeling process for its realistic deformation and real-time rendering. As P4 mentioned, \textit{"Usually, I rarely look at the texture rendering while editing shapes because it’s very inefficient in traditional software."} P2 described the process of editing the 3D model of the mermaid, saying, \textit{"It feels like the mermaid is really swimming in front of me."} P3 and P8 especially appreciated the feature to switch between different material types, which allowed them to edit the shape based on different deformation effects under varying plasticity types and material parameters. P9 favored the use of gravity during the modeling process to naturally stack objects together. Besides, the interactive feedback was another major advantage, allowing participants to immediately see the deformation results of their actions and make corresponding adjustments.

\paragraph{Encountered Problems and Suggested Features}
Our participants also identified a few areas for improvement. Hand-tracking instability was mentioned most frequently (P3, P6, P9, and P11), particularly when adjusting localized features of the object, resulting in unintended shape changes. Even though we have applied temporal smoothing to the hand movements, this limitation is largely due to the constraints of the hand-tracking hardware. Therefore, participants expressed the need for an undo feature to help them recover from mistakes. Meanwhile, P2, P9, and P11 also suggested reducing the degrees of freedom for moving or rotating objects during the modeling process, similar to the approach used in pottery making. P1 and P7 also noted occasional performance issues in high-resolution scenarios. For instance, P1 found that merging multiple objects together led to a loss of surface details. Besides, P1 and P8 suggested adding an operation to separate objects. P2 and P10 mentioned that objects would unintentionally collide with the simulation domain boundary during editing, and we have followed their suggestion of adding a one-click centering feature to avoid this issue.
\\

In summary, participants generally appreciated the system's efficiency in using hands, tools, or mid-air gestural input in their 3D modeling experience. While expert users acknowledged the continued necessity of conventional tools for intricate detailing, VR-Doh was particularly praised for quickly expressing design ideas, fostering creative exploration, and facilitating communication. While there were some concerns regarding precision and the lack of an undo feature, this did not overshadow the tool's potential for creating high-quality 3D models for both novice and expert users.

\subsection{US2: Comparison with Blender}
To further investigate the differences between VR-Doh and conventional desktop-based modeling software, e.g., Blender\footnote{\url{https://www.blender.org/}}, we invited six participants, including four with 3-7 years of modeling experience and two with no experience, to perform two goal-directed 3D modeling tasks.

\paragraph{Tasks \& Procedure}
Each participant was required to create both a Swiss roll and a donut using VR-Doh and Blender 4.3, guided by a step-by-step video tutorial. They were instructed to replicate the results demonstrated in the tutorial as closely as possible. When completing tasks with VR-Doh, participants watched the tutorial video until they claimed to have memorized the necessary operations. To minimize potential learning effects, the order of tools used was counterbalanced across participants. After completing the tasks, we conducted a semi-structured interview to invite participants to provide feedback and share their preferences regarding their modeling experiences with both tools.

\paragraph{Results}
We found that participants, on average, required less time to complete the Swiss roll (5.6 vs 9.4 min) and donut (5.2 vs 11.8 min) tasks using VR-Doh compared to Blender. Notably, one participant with modeling experience preferred using VR-Doh for the Swiss roll and Blender for the donut, indicating that VR-Doh may serve as a valuable complement to Blender in 3D modeling.

\paragraph{Key Differences}
Compared to Blender, we found VR-Doh provides a more natural modeling approach, making it easier for both experienced and novice participants to understand the operational steps for creating objects from the video. Four key advantages were identified in the user study:
\textit{1) Flexible selection:} Users can intuitively select and adjust editing areas on an object using dexterous hand movements. For example, pinching and twisting with the thumb and index finger can flatten surfaces or smooth protrusions, while shape deformation often aligns with hand gestures, enabling "what-you-see-is-what-you-get" precision. This also avoids over-selection or under-selection issues common in 2D-projected views of conventional tools.
\textit{2) Realistic deformation:} Unlike conventional methods such as ARAP, VR-Doh uses elastoplastic deformation, offering more realistic results. Users can edit overall shapes and fine details, stack objects using gravity and contact, and leverage material-specific properties for intuitive sculpting.
\textit{3) Pose editing without rigging:} Specific body parts can be selected and adjusted directly, allowing pose edits without the need for skeletal rigging. Joint smoothness is automatically handled by the elastoplastic simulation.
\textit{4) Physics-based cutting:} Objects can be separated at weak points by applying physical forces, such as breaking small connections, without the need to manually adjust cutting planes in complex geometries as in traditional 3D modeling software.
In summary, the above benefits of VR-Doh significantly enhance the efficiency of 3D modeling, enabling users to quickly figure out modeling steps and realize design ideas. As one participant noted, it helps avoid the common issue experienced with conventional modeling software, where "the hands cannot keep up with the mind."

While the main disadvantage was reported by our participants, in more specialized modeling tasks, the precision of 3D models created using VR-Doh was found to be relatively lower than that of Blender. This is primarily due to the limited number of particles available during real-time simulation and rendering. Additionally, the lack of haptic feedback often led to over-adjustments. Overall, we found that VR-Doh is better suited for shaping the overall structure of objects due to its intuitive operation, whereas Blender excels in detailed, localized operations because of its precise selection of points, edges, and faces. Therefore, the two tools exhibit a complementary relationship.

%% file: chapter08-discussion.tex
\section{Discussion}
\paragraph{Limitations}
Although VR-Doh improves the accessibility of physics-based 3D modeling through natural hand interactions, it has several limitations. The system is particularly effective for large-scale deformations, such as global stretching and bending, thanks to the realistic simulation of elastoplastic material behavior. However, fine-grained editing tasks, such as sculpting sharp creases or intricate surface details, remain challenging due to computational budget constraints. Additionally, vision-based hand tracking in consumer-grade VR headsets introduces minor positional jitter that affects precision during delicate manipulations. The absence of haptic feedback further complicates force estimation, especially in visually occluded regions, increasing the risk of unintended shape modifications.

\paragraph{Future Works}
To address precision limitations, future research can focus on hybrid interaction paradigms that blend the tactile engagement of physical sculpting with the precision and flexibility of digital tools~\cite{ma2024get}. The integration of haptic gloves could offer dual advantages: improving tracking stability through additional sensing modalities and enabling force feedback during precision tasks. Alternative feedback mechanisms such as auditory cues could also be explored. Such hardware advancements would further enhance our immersiveness by achieving higher operational accuracy while preserving the system's core strength in effectively handling large deformations.
Another meaningful direction involves developing efficient undo mechanisms that enable users to iteratively refine high-precision edits through trial and error. 
While our current implementation relies on periodic state saving for basic undo functionality, specialized compression schemes for incremental deformation states could enable more practical undo/redo operations without excessive memory overhead.

%% file: chapter09-conclusion.tex
\section{Conclusion}
We presented VR-Doh, a VR-based 3D modeling system that integrates advanced elastoplastic simulation with intuitive, hand-based interaction. Leveraging the Material Point Method (MPM) and incorporating innovations such as localized simulation, particle-level collision handling, and decoupled appearance and physical representations, the system achieves real-time responsiveness while delivering high-quality simulation and rendering. These technical innovations make VR-Doh accessible to novice users while empowering experts to perform detailed and expressive modeling. Extensive evaluations and user studies showcased the system's ability to provide an intuitive and immersive modeling experience. Compared to traditional tools, VR-Doh offers a more natural and accessible approach, with user-identified advantages such as flexible selection, realistic deformation, rig-free pose editing, and physics-based cutting, paving the way for a new paradigm in 3D modeling.

%% file: sigconf.bbl

\begin{thebibliography}{44}


\ifx \showCODEN    \undefined \def \showCODEN     #1{\unskip}     \fi
\ifx \showDOI      \undefined \def \showDOI       #1{#1}\fi
\ifx \showISBNx    \undefined \def \showISBNx     #1{\unskip}     \fi
\ifx \showISBNxiii \undefined \def \showISBNxiii  #1{\unskip}     \fi
\ifx \showISSN     \undefined \def \showISSN      #1{\unskip}     \fi
\ifx \showLCCN     \undefined \def \showLCCN      #1{\unskip}     \fi
\ifx \shownote     \undefined \def \shownote      #1{#1}          \fi
\ifx \showarticletitle \undefined \def \showarticletitle #1{#1}   \fi
\ifx \showURL      \undefined \def \showURL       {\relax}        \fi
\providecommand\bibfield[2]{#2}
\providecommand\bibinfo[2]{#2}
\providecommand\natexlab[1]{#1}
\providecommand\showeprint[2][]{arXiv:#2}

\bibitem[Arora et~al\mbox{.}(2019)]%
        {arora2019magicalhands}
\bibfield{author}{\bibinfo{person}{Rahul Arora}, \bibinfo{person}{Rubaiat~Habib Kazi}, \bibinfo{person}{Danny~M Kaufman}, \bibinfo{person}{Wilmot Li}, {and} \bibinfo{person}{Karan Singh}.} \bibinfo{year}{2019}\natexlab{}.
\newblock \showarticletitle{Magicalhands: Mid-air hand gestures for animating in vr}. In \bibinfo{booktitle}{\emph{Proceedings of the 32nd annual ACM symposium on user interface software and technology}}. \bibinfo{pages}{463--477}.
\newblock


\bibitem[Barreiro et~al\mbox{.}(2021)]%
        {barreiro2021natural}
\bibfield{author}{\bibinfo{person}{H{\'e}ctor Barreiro}, \bibinfo{person}{Joan Torres}, {and} \bibinfo{person}{Miguel~A Otaduy}.} \bibinfo{year}{2021}\natexlab{}.
\newblock \showarticletitle{Natural tactile interaction with virtual clay}. In \bibinfo{booktitle}{\emph{2021 IEEE World Haptics Conference (WHC)}}. IEEE, \bibinfo{pages}{403--408}.
\newblock


\bibitem[Chatterjee(2024)]%
        {chatterjee2024free}
\bibfield{author}{\bibinfo{person}{Shounak Chatterjee}.} \bibinfo{year}{2024}\natexlab{}.
\newblock \showarticletitle{Free-form Shape Modeling in XR: A Systematic Review}.
\newblock \bibinfo{journal}{\emph{arXiv preprint arXiv:2401.00924}} (\bibinfo{year}{2024}).
\newblock


\bibitem[Deng et~al\mbox{.}(2023)]%
        {deng2023phyvr}
\bibfield{author}{\bibinfo{person}{Hanchen Deng}, \bibinfo{person}{Jin Li}, \bibinfo{person}{Yang Gao}, \bibinfo{person}{Xiaohui Liang}, \bibinfo{person}{Hongyu Wu}, {and} \bibinfo{person}{Aimin Hao}.} \bibinfo{year}{2023}\natexlab{}.
\newblock \showarticletitle{PhyVR: Physics-based Multi-material and Free-hand Interaction in VR}. In \bibinfo{booktitle}{\emph{2023 IEEE International Symposium on Mixed and Augmented Reality (ISMAR)}}. IEEE, \bibinfo{pages}{454--462}.
\newblock


\bibitem[Dewaele and Cani(2004)]%
        {dewaele2004interactive}
\bibfield{author}{\bibinfo{person}{Guillaume Dewaele} {and} \bibinfo{person}{Marie-Paule Cani}.} \bibinfo{year}{2004}\natexlab{}.
\newblock \showarticletitle{Interactive global and local deformations for virtual clay}.
\newblock \bibinfo{journal}{\emph{Graphical Models}} \bibinfo{volume}{66}, \bibinfo{number}{6} (\bibinfo{year}{2004}), \bibinfo{pages}{352--369}.
\newblock


\bibitem[Eroglu et~al\mbox{.}(2018)]%
        {eroglu2018fluid}
\bibfield{author}{\bibinfo{person}{Sevinc Eroglu}, \bibinfo{person}{Sascha Gebhardt}, \bibinfo{person}{Patric Schmitz}, \bibinfo{person}{Dominik Rausch}, {and} \bibinfo{person}{Torsten~Wolfgang Kuhlen}.} \bibinfo{year}{2018}\natexlab{}.
\newblock \showarticletitle{Fluid sketching―Immersive sketching based on fluid flow}. In \bibinfo{booktitle}{\emph{2018 IEEE Conference on Virtual Reality and 3D User Interfaces (VR)}}. IEEE, \bibinfo{pages}{475--482}.
\newblock


\bibitem[Fang et~al\mbox{.}(2021)]%
        {Fang2021IDP}
\bibfield{author}{\bibinfo{person}{Yu Fang}, \bibinfo{person}{Minchen Li}, \bibinfo{person}{Chenfanfu Jiang}, {and} \bibinfo{person}{Danny~M. Kaufman}.} \bibinfo{year}{2021}\natexlab{}.
\newblock \showarticletitle{Guaranteed Globally Injective 3D Deformation Processing}.
\newblock \bibinfo{journal}{\emph{ACM Trans. Graph. (SIGGRAPH)}} \bibinfo{volume}{40}, \bibinfo{number}{4}, Article \bibinfo{articleno}{75} (\bibinfo{year}{2021}).
\newblock


\bibitem[Faraj et~al\mbox{.}(2013)]%
        {faraj2013progressive}
\bibfield{author}{\bibinfo{person}{Noura Faraj}, \bibinfo{person}{Jean-Marc Thiery}, {and} \bibinfo{person}{Tamy Boubekeur}.} \bibinfo{year}{2013}\natexlab{}.
\newblock \showarticletitle{Progressive medial axis filtration}.
\newblock In \bibinfo{booktitle}{\emph{SIGGRAPH Asia 2013 Technical Briefs}}. \bibinfo{pages}{1--4}.
\newblock


\bibitem[Galyean and Hughes(1991)]%
        {galyean1991sculpting}
\bibfield{author}{\bibinfo{person}{Tinsley~A Galyean} {and} \bibinfo{person}{John~F Hughes}.} \bibinfo{year}{1991}\natexlab{}.
\newblock \showarticletitle{Sculpting: An interactive volumetric modeling technique}.
\newblock \bibinfo{journal}{\emph{ACM SIGGRAPH Computer Graphics}} \bibinfo{volume}{25}, \bibinfo{number}{4} (\bibinfo{year}{1991}), \bibinfo{pages}{267--274}.
\newblock


\bibitem[H{\"o}ll et~al\mbox{.}(2018)]%
        {holl2018efficient}
\bibfield{author}{\bibinfo{person}{Markus H{\"o}ll}, \bibinfo{person}{Markus Oberweger}, \bibinfo{person}{Clemens Arth}, {and} \bibinfo{person}{Vincent Lepetit}.} \bibinfo{year}{2018}\natexlab{}.
\newblock \showarticletitle{Efficient physics-based implementation for realistic hand-object interaction in virtual reality}. In \bibinfo{booktitle}{\emph{2018 IEEE conference on virtual reality and 3D user interfaces (VR)}}. IEEE, \bibinfo{pages}{175--182}.
\newblock


\bibitem[Hu et~al\mbox{.}(2018)]%
        {hu2018moving}
\bibfield{author}{\bibinfo{person}{Yuanming Hu}, \bibinfo{person}{Yu Fang}, \bibinfo{person}{Ziheng Ge}, \bibinfo{person}{Ziyin Qu}, \bibinfo{person}{Yixin Zhu}, \bibinfo{person}{Andre Pradhana}, {and} \bibinfo{person}{Chenfanfu Jiang}.} \bibinfo{year}{2018}\natexlab{}.
\newblock \showarticletitle{A moving least squares material point method with displacement discontinuity and two-way rigid body coupling}.
\newblock \bibinfo{journal}{\emph{ACM Transactions on Graphics (TOG)}} \bibinfo{volume}{37}, \bibinfo{number}{4} (\bibinfo{year}{2018}), \bibinfo{pages}{1--14}.
\newblock


\bibitem[Hu et~al\mbox{.}(2019)]%
        {hu2019taichi}
\bibfield{author}{\bibinfo{person}{Yuanming Hu}, \bibinfo{person}{Tzu-Mao Li}, \bibinfo{person}{Luke Anderson}, \bibinfo{person}{Jonathan Ragan-Kelley}, {and} \bibinfo{person}{Fr{\'e}do Durand}.} \bibinfo{year}{2019}\natexlab{}.
\newblock \showarticletitle{Taichi: a language for high-performance computation on spatially sparse data structures}.
\newblock \bibinfo{journal}{\emph{ACM Transactions on Graphics (TOG)}} \bibinfo{volume}{38}, \bibinfo{number}{6} (\bibinfo{year}{2019}), \bibinfo{pages}{1--16}.
\newblock


\bibitem[Jacobs and Froehlich(2011)]%
        {jacobs2011soft}
\bibfield{author}{\bibinfo{person}{Jan Jacobs} {and} \bibinfo{person}{Bernd Froehlich}.} \bibinfo{year}{2011}\natexlab{}.
\newblock \showarticletitle{A soft hand model for physically-based manipulation of virtual objects}. In \bibinfo{booktitle}{\emph{2011 IEEE virtual reality conference}}. IEEE, \bibinfo{pages}{11--18}.
\newblock


\bibitem[Jang et~al\mbox{.}(2014)]%
        {jang2014airsculpt}
\bibfield{author}{\bibinfo{person}{Sung-A Jang}, \bibinfo{person}{Hyung-il Kim}, \bibinfo{person}{Woontack Woo}, {and} \bibinfo{person}{Graham Wakefield}.} \bibinfo{year}{2014}\natexlab{}.
\newblock \showarticletitle{Airsculpt: A wearable augmented reality 3d sculpting system}. In \bibinfo{booktitle}{\emph{Distributed, Ambient, and Pervasive Interactions: Second International Conference, DAPI 2014, Held as Part of HCI Interational 2014, Heraklion, Crete, Greece, June 22-27, 2014. Proceedings 2}}. Springer, \bibinfo{pages}{130--141}.
\newblock


\bibitem[Jiang et~al\mbox{.}(2016)]%
        {jiang2016material}
\bibfield{author}{\bibinfo{person}{Chenfanfu Jiang}, \bibinfo{person}{Craig Schroeder}, \bibinfo{person}{Joseph Teran}, \bibinfo{person}{Alexey Stomakhin}, {and} \bibinfo{person}{Andrew Selle}.} \bibinfo{year}{2016}\natexlab{}.
\newblock \showarticletitle{The material point method for simulating continuum materials}.
\newblock In \bibinfo{booktitle}{\emph{Acm siggraph 2016 courses}}. \bibinfo{pages}{1--52}.
\newblock


\bibitem[Jiang et~al\mbox{.}(2024)]%
        {jiang2024vr}
\bibfield{author}{\bibinfo{person}{Ying Jiang}, \bibinfo{person}{Chang Yu}, \bibinfo{person}{Tianyi Xie}, \bibinfo{person}{Xuan Li}, \bibinfo{person}{Yutao Feng}, \bibinfo{person}{Huamin Wang}, \bibinfo{person}{Minchen Li}, \bibinfo{person}{Henry Lau}, \bibinfo{person}{Feng Gao}, \bibinfo{person}{Yin Yang}, {and} \bibinfo{person}{Chenfanfu Jiang}.} \bibinfo{year}{2024}\natexlab{}.
\newblock \showarticletitle{VR-GS: a physical dynamics-aware interactive gaussian splatting system in virtual reality}. In \bibinfo{booktitle}{\emph{ACM SIGGRAPH 2024 Conference Papers}}. \bibinfo{pages}{1--1}.
\newblock


\bibitem[Keefe et~al\mbox{.}(2008)]%
        {keefe2008tech}
\bibfield{author}{\bibinfo{person}{Daniel~F Keefe}, \bibinfo{person}{Robert~C Zeleznik}, {and} \bibinfo{person}{David~H Laidlaw}.} \bibinfo{year}{2008}\natexlab{}.
\newblock \showarticletitle{Tech-note: Dynamic dragging for input of 3D trajectories}. In \bibinfo{booktitle}{\emph{2008 IEEE Symposium on 3D User Interfaces}}. IEEE, \bibinfo{pages}{51--54}.
\newblock


\bibitem[Kennedy et~al\mbox{.}(1993)]%
        {kennedy1993simulator}
\bibfield{author}{\bibinfo{person}{Robert~S Kennedy}, \bibinfo{person}{Norman~E Lane}, \bibinfo{person}{Kevin~S Berbaum}, {and} \bibinfo{person}{Michael~G Lilienthal}.} \bibinfo{year}{1993}\natexlab{}.
\newblock \showarticletitle{Simulator sickness questionnaire: An enhanced method for quantifying simulator sickness}.
\newblock \bibinfo{journal}{\emph{The international journal of aviation psychology}} \bibinfo{volume}{3}, \bibinfo{number}{3} (\bibinfo{year}{1993}), \bibinfo{pages}{203--220}.
\newblock


\bibitem[Kerbl et~al\mbox{.}(2023)]%
        {kerbl20233d}
\bibfield{author}{\bibinfo{person}{Bernhard Kerbl}, \bibinfo{person}{Georgios Kopanas}, \bibinfo{person}{Thomas Leimk{\"u}hler}, {and} \bibinfo{person}{George Drettakis}.} \bibinfo{year}{2023}\natexlab{}.
\newblock \showarticletitle{3D Gaussian Splatting for Real-Time Radiance Field Rendering.}
\newblock \bibinfo{journal}{\emph{ACM Trans. Graph.}} \bibinfo{volume}{42}, \bibinfo{number}{4} (\bibinfo{year}{2023}), \bibinfo{pages}{139--1}.
\newblock


\bibitem[Kingma and Ba(2014)]%
        {Kingma2014AdamAM}
\bibfield{author}{\bibinfo{person}{Diederik~P. Kingma} {and} \bibinfo{person}{Jimmy Ba}.} \bibinfo{year}{2014}\natexlab{}.
\newblock \showarticletitle{Adam: A Method for Stochastic Optimization}.
\newblock \bibinfo{journal}{\emph{CoRR}}  \bibinfo{volume}{abs/1412.6980} (\bibinfo{year}{2014}).
\newblock
\urldef\tempurl%
\url{https://api.semanticscholar.org/CorpusID:6628106}
\showURL{%
\tempurl}


\bibitem[Kumar and Todorov(2015)]%
        {kumar2015mujoco}
\bibfield{author}{\bibinfo{person}{Vikash Kumar} {and} \bibinfo{person}{Emanuel Todorov}.} \bibinfo{year}{2015}\natexlab{}.
\newblock \showarticletitle{Mujoco haptix: A virtual reality system for hand manipulation}. In \bibinfo{booktitle}{\emph{2015 IEEE-RAS 15th International Conference on Humanoid Robots (Humanoids)}}. IEEE, \bibinfo{pages}{657--663}.
\newblock


\bibitem[Lan et~al\mbox{.}(2021)]%
        {lan2021medial}
\bibfield{author}{\bibinfo{person}{Lei Lan}, \bibinfo{person}{Yin Yang}, \bibinfo{person}{Danny Kaufman}, \bibinfo{person}{Junfeng Yao}, \bibinfo{person}{Minchen Li}, {and} \bibinfo{person}{Chenfanfu Jiang}.} \bibinfo{year}{2021}\natexlab{}.
\newblock \showarticletitle{Medial IPC: accelerated incremental potential contact with medial elastics}.
\newblock \bibinfo{journal}{\emph{ACM Transactions on Graphics}} \bibinfo{volume}{40}, \bibinfo{number}{4} (\bibinfo{year}{2021}).
\newblock


\bibitem[Li et~al\mbox{.}(2015)]%
        {li2015q}
\bibfield{author}{\bibinfo{person}{Pan Li}, \bibinfo{person}{Bin Wang}, \bibinfo{person}{Feng Sun}, \bibinfo{person}{Xiaohu Guo}, \bibinfo{person}{Caiming Zhang}, {and} \bibinfo{person}{Wenping Wang}.} \bibinfo{year}{2015}\natexlab{}.
\newblock \showarticletitle{Q-mat: Computing medial axis transform by quadratic error minimization}.
\newblock \bibinfo{journal}{\emph{ACM Transactions on Graphics (TOG)}} \bibinfo{volume}{35}, \bibinfo{number}{1} (\bibinfo{year}{2015}), \bibinfo{pages}{1--16}.
\newblock


\bibitem[Lougiakis et~al\mbox{.}(2024)]%
        {lougiakis2024comparing}
\bibfield{author}{\bibinfo{person}{Christos Lougiakis}, \bibinfo{person}{Jorge~Juan González}, \bibinfo{person}{Giorgos Ganias}, \bibinfo{person}{Akrivi Katifori}, \bibinfo{person}{Ioannis-Panagiotis}, {and} \bibinfo{person}{Maria Roussou}.} \bibinfo{year}{2024}\natexlab{}.
\newblock \showarticletitle{Comparing Physics-based Hand Interaction in Virtual Reality: Custom Soft Body Simulation vs. Off-the-Shelf Integrated Solution}. In \bibinfo{booktitle}{\emph{2024 IEEE Conference Virtual Reality and 3D User Interfaces (VR)}}. IEEE, \bibinfo{pages}{743--753}.
\newblock


\bibitem[Lund(2001)]%
        {lund2001measuring}
\bibfield{author}{\bibinfo{person}{Arnold~M Lund}.} \bibinfo{year}{2001}\natexlab{}.
\newblock \showarticletitle{Measuring usability with the use questionnaire12}.
\newblock \bibinfo{journal}{\emph{Usability interface}} \bibinfo{volume}{8}, \bibinfo{number}{2} (\bibinfo{year}{2001}), \bibinfo{pages}{3--6}.
\newblock


\bibitem[Ma et~al\mbox{.}(2024)]%
        {ma2024get}
\bibfield{author}{\bibinfo{person}{Xinyu Ma}, \bibinfo{person}{Zhe Yan}, \bibinfo{person}{Hengyu Meng}, \bibinfo{person}{Zhihao Cao}, \bibinfo{person}{Yanan Jin}, {and} \bibinfo{person}{Zeyu Wang}.} \bibinfo{year}{2024}\natexlab{}.
\newblock \showarticletitle{Get Your Hands Dirty? A Comparative Study of Tool Usage and Perceptual Engagement in Physical and Digital Sculpting}. In \bibinfo{booktitle}{\emph{Proceedings of the 16th Conference on Creativity \& Cognition}}. \bibinfo{pages}{358--373}.
\newblock


\bibitem[Marner and Thomas(2010)]%
        {marner2010augmented}
\bibfield{author}{\bibinfo{person}{Michael~R Marner} {and} \bibinfo{person}{Bruce~H Thomas}.} \bibinfo{year}{2010}\natexlab{}.
\newblock \showarticletitle{Augmented foam sculpting for capturing 3D models}. In \bibinfo{booktitle}{\emph{2010 IEEE Symposium on 3D User Interfaces (3DUI)}}. IEEE, \bibinfo{pages}{63--70}.
\newblock


\bibitem[McDonnell et~al\mbox{.}(2001)]%
        {mcdonnell2001virtual}
\bibfield{author}{\bibinfo{person}{Kevin~T McDonnell}, \bibinfo{person}{Hong Qin}, {and} \bibinfo{person}{Robert~A Wlodarczyk}.} \bibinfo{year}{2001}\natexlab{}.
\newblock \showarticletitle{Virtual clay: A real-time sculpting system with haptic toolkits}. In \bibinfo{booktitle}{\emph{Proceedings of the 2001 symposium on Interactive 3D graphics}}. \bibinfo{pages}{179--190}.
\newblock


\bibitem[Moo-Young et~al\mbox{.}(2021)]%
        {moo2021virtual}
\bibfield{author}{\bibinfo{person}{Joss~Kingdom Moo-Young}, \bibinfo{person}{Andrew Hogue}, {and} \bibinfo{person}{Veronika Szkudlarek}.} \bibinfo{year}{2021}\natexlab{}.
\newblock \showarticletitle{Virtual materiality: Realistic clay sculpting in vr}. In \bibinfo{booktitle}{\emph{Extended Abstracts of the 2021 Annual Symposium on Computer-Human Interaction in Play}}. \bibinfo{pages}{105--110}.
\newblock


\bibitem[Nealen et~al\mbox{.}(2007)]%
        {nealen2007fibermesh}
\bibfield{author}{\bibinfo{person}{Andrew Nealen}, \bibinfo{person}{Takeo Igarashi}, \bibinfo{person}{Olga Sorkine}, {and} \bibinfo{person}{Marc Alexa}.} \bibinfo{year}{2007}\natexlab{}.
\newblock \showarticletitle{Fibermesh: designing freeform surfaces with 3d curves}.
\newblock In \bibinfo{booktitle}{\emph{ACM SIGGRAPH 2007 papers}}. \bibinfo{pages}{41--es}.
\newblock


\bibitem[Peng et~al\mbox{.}(2020)]%
        {peng2020autocomplete}
\bibfield{author}{\bibinfo{person}{Mengqi Peng}, \bibinfo{person}{Li-yi Wei}, \bibinfo{person}{Rubaiat~Habib Kazi}, {and} \bibinfo{person}{Vladimir~G Kim}.} \bibinfo{year}{2020}\natexlab{}.
\newblock \showarticletitle{Autocomplete animated sculpting}. In \bibinfo{booktitle}{\emph{Proceedings of the 33rd Annual ACM Symposium on User Interface Software and Technology}}. \bibinfo{pages}{760--777}.
\newblock


\bibitem[Pihuit et~al\mbox{.}(2008)]%
        {pihuit2008hands}
\bibfield{author}{\bibinfo{person}{Adeline Pihuit}, \bibinfo{person}{Paul~G Kry}, {and} \bibinfo{person}{Marie-Paule Cani}.} \bibinfo{year}{2008}\natexlab{}.
\newblock \showarticletitle{Hands on virtual clay}. In \bibinfo{booktitle}{\emph{2008 IEEE International Conference on Shape Modeling and Applications}}. IEEE, \bibinfo{pages}{267--268}.
\newblock


\bibitem[Pohl and Mottelson(2022)]%
        {Pohl2022Hands}
\bibfield{author}{\bibinfo{person}{Henning Pohl} {and} \bibinfo{person}{Aske Mottelson}.} \bibinfo{year}{2022}\natexlab{}.
\newblock \showarticletitle{Hafnia Hands: A Multi-Skin Hand Texture Resource for Virtual Reality Research}.
\newblock \bibinfo{journal}{\emph{Frontiers in Virtual Reality}}  \bibinfo{volume}{3} (\bibinfo{year}{2022}).
\newblock
\showISSN{2673-4192}


\bibitem[Rosales et~al\mbox{.}(2021)]%
        {rosales2021adaptibrush}
\bibfield{author}{\bibinfo{person}{Enrique Rosales}, \bibinfo{person}{Chrystiano Ara{\'u}jo}, \bibinfo{person}{Jafet Rodriguez}, \bibinfo{person}{Nicholas Vining}, \bibinfo{person}{Dongwook Yoon}, {and} \bibinfo{person}{Alla Sheffer}.} \bibinfo{year}{2021}\natexlab{}.
\newblock \showarticletitle{AdaptiBrush: adaptive general and predictable VR ribbon brush.}
\newblock \bibinfo{journal}{\emph{ACM Trans. Graph.}} \bibinfo{volume}{40}, \bibinfo{number}{6} (\bibinfo{year}{2021}), \bibinfo{pages}{247--1}.
\newblock


\bibitem[Rosales et~al\mbox{.}(2019)]%
        {rosales2019surfacebrush}
\bibfield{author}{\bibinfo{person}{Enrique Rosales}, \bibinfo{person}{Jafet Rodriguez}, {and} \bibinfo{person}{Alla Sheffer}.} \bibinfo{year}{2019}\natexlab{}.
\newblock \showarticletitle{SurfaceBrush: from virtual reality drawings to manifold surfaces}.
\newblock \bibinfo{journal}{\emph{arXiv preprint arXiv:1904.12297}} (\bibinfo{year}{2019}).
\newblock


\bibitem[Schkolne et~al\mbox{.}(2001)]%
        {schkolne2001surface}
\bibfield{author}{\bibinfo{person}{Steven Schkolne}, \bibinfo{person}{Michael Pruett}, {and} \bibinfo{person}{Peter Schr{\"o}der}.} \bibinfo{year}{2001}\natexlab{}.
\newblock \showarticletitle{Surface drawing: creating organic 3D shapes with the hand and tangible tools}. In \bibinfo{booktitle}{\emph{Proceedings of the SIGCHI conference on Human factors in computing systems}}. \bibinfo{pages}{261--268}.
\newblock


\bibitem[Schulz et~al\mbox{.}(2019)]%
        {schulz2019rodmesh}
\bibfield{author}{\bibinfo{person}{HJ Schulz}, \bibinfo{person}{M Teschner}, {and} \bibinfo{person}{M Wimmer}.} \bibinfo{year}{2019}\natexlab{}.
\newblock \showarticletitle{Rodmesh: Two-handed 3D surface modeling in virtual reality}. In \bibinfo{booktitle}{\emph{Vision, modeling and visualization}}. \bibinfo{pages}{1--10}.
\newblock


\bibitem[Sheng et~al\mbox{.}(2006)]%
        {sheng2006interface}
\bibfield{author}{\bibinfo{person}{Jia Sheng}, \bibinfo{person}{Ravin Balakrishnan}, {and} \bibinfo{person}{Karan Singh}.} \bibinfo{year}{2006}\natexlab{}.
\newblock \showarticletitle{An interface for virtual 3D sculpting via physical proxy.}. In \bibinfo{booktitle}{\emph{GRAPHITE}}, Vol.~\bibinfo{volume}{6}. \bibinfo{pages}{213--220}.
\newblock


\bibitem[Smith et~al\mbox{.}(2020)]%
        {smith2020constraining}
\bibfield{author}{\bibinfo{person}{Breannan Smith}, \bibinfo{person}{Chenglei Wu}, \bibinfo{person}{He Wen}, \bibinfo{person}{Patrick Peluse}, \bibinfo{person}{Yaser Sheikh}, \bibinfo{person}{Jessica~K Hodgins}, {and} \bibinfo{person}{Takaaki Shiratori}.} \bibinfo{year}{2020}\natexlab{}.
\newblock \showarticletitle{Constraining dense hand surface tracking with elasticity}.
\newblock \bibinfo{journal}{\emph{ACM Transactions on Graphics (ToG)}} \bibinfo{volume}{39}, \bibinfo{number}{6} (\bibinfo{year}{2020}), \bibinfo{pages}{1--14}.
\newblock


\bibitem[Stolpner et~al\mbox{.}(2011)]%
        {stolpner2011medial}
\bibfield{author}{\bibinfo{person}{Svetlana Stolpner}, \bibinfo{person}{Paul Kry}, {and} \bibinfo{person}{Kaleem Siddiqi}.} \bibinfo{year}{2011}\natexlab{}.
\newblock \showarticletitle{Medial spheres for shape approximation}.
\newblock \bibinfo{journal}{\emph{IEEE transactions on pattern analysis and machine intelligence}} \bibinfo{volume}{34}, \bibinfo{number}{6} (\bibinfo{year}{2011}), \bibinfo{pages}{1234--1240}.
\newblock


\bibitem[Sun et~al\mbox{.}(2015)]%
        {sun2015medial}
\bibfield{author}{\bibinfo{person}{Feng Sun}, \bibinfo{person}{Yi-King Choi}, \bibinfo{person}{Yizhou Yu}, {and} \bibinfo{person}{Wenping Wang}.} \bibinfo{year}{2015}\natexlab{}.
\newblock \showarticletitle{Medial meshes--a compact and accurate representation of medial axis transform}.
\newblock \bibinfo{journal}{\emph{IEEE transactions on visualization and computer graphics}} \bibinfo{volume}{22}, \bibinfo{number}{3} (\bibinfo{year}{2015}), \bibinfo{pages}{1278--1290}.
\newblock


\bibitem[Verschoor et~al\mbox{.}(2018)]%
        {verschoor2018soft}
\bibfield{author}{\bibinfo{person}{Mickeal Verschoor}, \bibinfo{person}{Daniel Lobo}, {and} \bibinfo{person}{Miguel~A Otaduy}.} \bibinfo{year}{2018}\natexlab{}.
\newblock \showarticletitle{Soft hand simulation for smooth and robust natural interaction}. In \bibinfo{booktitle}{\emph{2018 IEEE conference on virtual reality and 3D user interfaces (VR)}}. IEEE, \bibinfo{pages}{183--190}.
\newblock


\bibitem[Xie et~al\mbox{.}(2024)]%
        {xie2024physgaussian}
\bibfield{author}{\bibinfo{person}{Tianyi Xie}, \bibinfo{person}{Zeshun Zong}, \bibinfo{person}{Yuxing Qiu}, \bibinfo{person}{Xuan Li}, \bibinfo{person}{Yutao Feng}, \bibinfo{person}{Yin Yang}, {and} \bibinfo{person}{Chenfanfu Jiang}.} \bibinfo{year}{2024}\natexlab{}.
\newblock \showarticletitle{Physgaussian: Physics-integrated 3d gaussians for generative dynamics}. In \bibinfo{booktitle}{\emph{Proceedings of the IEEE/CVF Conference on Computer Vision and Pattern Recognition}}. \bibinfo{pages}{4389--4398}.
\newblock


\bibitem[Yu et~al\mbox{.}(2021)]%
        {yu2021cassie}
\bibfield{author}{\bibinfo{person}{Emilie Yu}, \bibinfo{person}{Rahul Arora}, \bibinfo{person}{Tibor Stanko}, \bibinfo{person}{J~Andreas B{\ae}rentzen}, \bibinfo{person}{Karan Singh}, {and} \bibinfo{person}{Adrien Bousseau}.} \bibinfo{year}{2021}\natexlab{}.
\newblock \showarticletitle{Cassie: Curve and surface sketching in immersive environments}. In \bibinfo{booktitle}{\emph{Proceedings of the 2021 CHI Conference on Human Factors in Computing Systems}}. \bibinfo{pages}{1--14}.
\newblock


\end{thebibliography}
